\documentclass[journal]{IEEEtran}
\ifCLASSINFOpdf
\else
\fi
\usepackage{xcolor}

\usepackage{cite}
\usepackage{algpseudocode}
\usepackage{amsmath,amssymb,amsfonts}
\usepackage{graphicx}
\usepackage{epstopdf}

\newcommand*{\Scale}[2][4]{\scalebox{#1}{$#2$}}%

\newtheorem{thm}{Theorem}

\usepackage{textcomp}

\begin{document}
%

\title{Data-driven Dynamic Multi-objective Optimal Control: An  Aspiration-satisfying Reinforcement Learning Approach}

%

\author{Majid~Mazouchi,~\IEEEmembership{Member,~IEEE,}
        Yongliang~Yang,~\IEEEmembership{Member,~IEEE,}
        and~Hamidreza~Modares,~\IEEEmembership{Senior ~Member,~IEEE}

\thanks{M. Mazouchi and H. Modares are with Michigan State University, East Lansing, MI 48824 USA e-mail: {Mazouchi, Modaresh}@msu.edu.}
\thanks{Yongliang Yang is with Key Laboratory of Knowledge Automation for Industrial Processes of Ministry of Education, School of Automation and Electrical Engineering, University of Science and Technology Beijing, Beijing 10083, China e-mail: {yangyongliang}@ieee.org.}
}

%
%

\markboth{}%
{Shell \MakeLowercase{\textit{et al.}}: Data-driven dynamic multi-objective optimal control: A Hamiltonian-inequality driven satisficing reinforcement learning approach}
%



\maketitle

\begin{abstract}
This paper presents an iterative data-driven algorithm for solving dynamic multi-objective (MO) optimal control problems arising in control of nonlinear continuous-time systems. It is first shown that the Hamiltonian functional corresponding to each objective can be leveraged to compare the performance of admissible policies. Hamiltonian-inequalities are then used for which their satisfaction guarantees satisfying the objectives' aspirations. 
{Relaxed Hamilton-Jacobi-bellman (HJB) equations in terms of HJB inequalities are then solved in a dynamic  constrained MO framework to find Pareto-optimal solutions.} Relation to satisficing (good enough) decision-making framework is shown. A Sum-of-Square (SOS) based iterative algorithm is developed to solve the formulated aspiration-satisfying MO optimization. To obviate the requirement of complete knowledge of the system dynamics, a data-driven satisficing reinforcement learning approach is proposed to solve the SOS optimization problem in real-time using only the information of the system trajectories measured during a time interval without having full knowledge of the system dynamics. Finally, two simulation examples are utilized to verify the analytical results of the proposed algorithm.

\end{abstract}

\begin{IEEEkeywords}
Multi-objective optimization, Reinforcement learning, Saticficing control, Sum-of-squares program.
\end{IEEEkeywords}

%
\IEEEpeerreviewmaketitle

\section{Introduction}
\IEEEPARstart{I}{n} most of the real-world control applications such as autonomous vehicles, the system designer must account for multiple objectives (such as safety, control effort, transient performance, comfort, etc.) to evaluate candidate control policies. However, since there usually exist conflicts between objectives and the objectives' preferences might change over time, a control policy is best realized by finding an appropriate context-dependent trade-off among objectives. A multi-objective (MO) optimal control framework that trades-off among objectives and could explicitly account for objectives' aspirations must be devised to deal with this issue. 

While MO optimization has been widely utilized to find a diverse set of efficient solutions (see for examples \cite{Toivonen:1986}, \cite{Toivonen:1989}, \cite{Marler:2004}, \cite{Gambier:2011}, \cite{Peitz:2018}, and \cite{Roijers:2013}), there are at least three challenges in control of dynamical systems with multiple objectives that are not well addressed. First, most of the existing MO optimization frameworks assume that the objective functions to be optimized are static which solves one-shot decision making problems. {In the control engineering systems, however, several objectives must be optimized over a horizon  \cite{Logist:2010}, \cite{Ober-Blobaum:2012}, and performing a sequence of static one-shot optimizations results in myopic short-sighted decisions that might lead to infeasibility in face of uncertainties.} Second, to successfully operate in a changing and uncertain environment, systems such as self-driving cars must learn multiple potential solutions for different situational objectives and apply autonomously the one with the appropriate trade-off as the situation becomes apparent. While solving several optimal control problems for a diverse set of preferences using a weighted sum of objectives can produce diverse solutions, however, since different objectives have different physical meanings and units, their scales are incomparable and the weighted-sum approach cannot capture the aspiration level (i.e., level of satisfaction) of each objective function for each context. Moreover, these methods cannot learn control policies in the nonconvex parts of the Pareto optimal set \cite{Caramia:2008}, \cite{Das:1997}. Finally, the uncertainty of the system's dynamics must also be taken into account when optimizing multiple objectives. This is mainly ignored in the existing approaches. 

Reinforcement Learning (RL) has been widely used to find optimal controllers for systems with uncertain dynamics. Most of the existing RL algorithms are presented for single-objective optimal control problems \cite{Lewis:2009}, \cite{Vamvoudakis:2010},  \cite{Jiang:2015}, \cite{Wang:2014}, \cite{Kamalapurkar:2018}, \cite{Modares:2016},
\cite{Beuchat:2020}, \cite{Tanzanakis:2020}. Recently, there has been a surge of interest in the study of MO reinforcement learning (MORL) problems  \cite{Logist:2010}, \cite{Caramia:2008}, \cite{Kang:2004}, \cite{Lopez:2019}, and \cite{Moffaert:2014}. Nevertheless, most of the existing MORL algorithms assume a given preference and find a single best policy corresponding to it based on the weighted sum of the objective functions. {Although solving several optimal control problems for a diverse set of preferences using a weighted sum of objectives can produce diverse solutions (see for example \cite{Barrett:2008}-\cite{Chen:2019}), these methods cannot learn control policies in the nonconvex parts of the Pareto optimal set \cite{Caramia:2008},\cite{Das:1997}. Moreover, since different objectives have different physical meanings and units, their scales are incomparable and the weighted-sum approach cannot capture the aspiration level (i.e., level of satisfaction) of each objective function for each context. Finally, these existing results are developed for Markov Decision Processes (MDPs), are offline methods, and need complete system dynamics' knowledge.}  {An online adaptive value iteration learning algorithm is developed in \cite{Abouheaf:2019} to solve a MO tracking control problem for a special class of discrete-time nonlinear systems, assuming that there is no conflict among objectives.}

 {This paper presents an alternative framework and solution approach to the multi-objective optimal control problem. In terms of framework, inspired by the $\epsilon$-constrained approach for static multi-objective optimization, only the main objective is optimized, and aspiration-satisfying constraints are imposed on other objectives that determine their level of satisfaction. This approach will relax solving HJB equalities required in standard dynamic MO problems (e.g., [20]) to solving HJB inequalities that are much more powerful in resolving conflicts between objectives. In terms of the solution approach, a novel satisficing reinforcement learning method is presented to learn the solution to the formulated problem without requiring the knowledge of the system dynamics and using only measured data along the system’s trajectories. Moreover, it is much easier to incorporate prior knowledge of preferences in terms of aspirations than the weight of objective functions (as performed in most MO RL literature), as the weighted-sum approach does cannot evaluate each objective function separately and cannot, therefore, incorporate any prior knowledge about expected performance for each objective.}  
 
 {More specifically, in this paper, it is first shown that the Hamiltonian functional corresponding to each objective can be leveraged to compare the performance of admissible policies. It is also shown that the aspiration level of each objective (i.e., the level of the performance at which the objective is satisfactory) can be imposed using a Hamiltonian inequality approach. It is shown that optimizing one objective while imposing an aspiration level on other objectives provides a Pareto solution. A set of solutions can then be found by imposing allowable aspiration on objective functions. For the solution approach, a data-driven Sum-of-Squares (SOS) based iterative algorithm is then developed to find a finite number of solutions of MO optimal control problems using only the information of the system trajectories measured during a time interval online in real-time. A simulation example is provided to verify the effectiveness of the presented approach.}

\textbf{Notations:} The following notations are needed throughout the paper. Let ${\Re ^n}$ and ${\Re ^{n \times m}}$  denote the $n$ dimensional real vector space, and the $n \times m$  real matrix space, respectively. Let $\mathbb{Z}^+$ and  ${\Re ^ + }$ denote the sets of all positive integers and real numbers, respectively. The set of all continuously differentiable polynomial functions is denoted by ${C^1}$. ${\cal P}$ denotes the set of all positive definite and proper polynomial functions in ${C^1}$.  {$\{y\}_j$ denotes the $j$-th element of vector $y$.} Let ${0_k} \in {\Re ^k}$  be the vector with all zeros and  ${1_k} \in {\Re ^k}$ the vector with all ones. Assume that ${y^1},{y^2} \in {\Re ^m}$. Then, ${y^1} \le {y^2}$ denotes weak componentwise order which implies $y_k^1 \le y_k^2$, $k = 1,...,m$. ${y^1} \prec {y^2}$ denotes Pareto order, which implies $y_k^1 \le y_k^2$, $k = 1,...,m$, ${y^1} \ne {y^2}$. ${y^1}\nprec{y^2}$ denotes that ${y^1}$ is not Pareto dominated by ${y^2}$. Assume that ${d_1},{d_2} \in \mathbb{Z}^+$, and ${d_2}{\rm{ }} \ge {\rm{ }}{d_1}$. Let $\Scale[0.9]{{  \theta : = \scriptstyle \left( {\begin{array}{*{20}{c}}
{n + {d_2}}\\
{{d_2}}
\end{array}} \right) - \scriptstyle \left( {\begin{array}{*{20}{c}}
{n + {d_1} - 1}\\
{{d_1} - 1}
\end{array}} \right)}}$. Then $ \overrightarrow m^{({d_1},{d_2})}(x) \in {\Re ^{\theta n}}$ is the arranged in lexicographic order vector of distinct monic monomials in terms of $x \in {\Re ^n}$ with degree $\kappa $  where ${d_1} \le \kappa  \le {d_2}$. Moreover, the set of all polynomials in $x \in {\Re ^n}$ with degree $\kappa $ is denoted by ${\cal R}{\left[ x \right]_{{d_1},{d_2}}}$.

%
%
%
%

 

\section{Problem Formulation}
Consider the following continuous-time nonlinear system 
\begin{align} 
\dot x = f(x) + g(x)u \label{eq:1}
\end{align}
where $x \in {\Re ^n}$ and $u \in {\Re ^m}$ are the state and control input of the system, respectively. In this work, we assume that $f(.):{\Re ^n} \to {\Re ^n}$  and $g(.):{\Re ^n} \to {\Re ^{n \times m}}$  are polynomial mappings and $f(0) = 0$. 

For simplicity, throughout the paper, we assume the system has only two objectives. The proposed approach, however, can be readily extended to more than two objectives. The two cost or objective functions associated with the system (\ref{eq:1}) are defined as 
\begin{align}
\Scale[1]{{J_i}\left( {x,u} \right) = \int\limits_0^\infty  {{r_i}\left( {x\left( t \right),u(x)} \right)dt} {\rm{ , }}i = 1,2,}  \label{eq:2}
\end{align}
{where ${r_i}(x,u) = {Q_i}(x) + {u^T}{R_i}u$, ${Q_i}(x)$ are the penalty functions on the states, and ${R_i} \in {\Re ^{m \times m}}$ are the symmetric penalty weighted matrices on the control efforts. In this paper, we require ${Q_i}(x)$ and ${R_i}$, $i = 1,2$ to be positive definite, so that the cost functions are well defined. Note that ${Q_i}(x)$ and ${R_i}$, $i = 1,2$ can be selected based on minimum-energy  considerations, minimum fatigue, etc.}



 {\textbf{Definition 1}. A control policy $u(x)$ is said to be admissible with respect to the cost functions ${J_i}(.)$, $i = 1,2$, if it is continuous, $u(0) = 0$, and it stabilizes the dynamics (\ref{eq:1}) and makes ${J_i(.)}$, $i = 1,2$ finite. The set of admissible policies is denoted by $\Phi $ in this paper.}

 \textbf{Definition 2}. The vector composed of the ideal values of the objective functions (\ref{eq:2}), denoted by ${J^{utopia}}: = {[J_1^{utopia},J_2^{utopia}]^T}$, is a utopia cost vector with respect to the system (\ref{eq:1}) and objective functions (\ref{eq:2}), if and only if
\begin{align}
J_i^{utopia} &= \mathop {\min \,}\limits_u {J_i}\left( {x,u} \right), i = 1,2. \label{eq:03} \\
&\quad s.t.\,\,\dot x = f(x) + g(x)u
\end{align}

{\textbf{Definition 3}. For the system (\ref{eq:1}) with two objectives given by (\ref{eq:2}), a  control policy ${u^1} \in \Phi $, is said to dominate a control policy  ${u^2} \in \Phi $, in Pareto sense, if and only if ${J_i}({u^1}) \le {J_i}({u^2})$, $\forall i \in \{ 1,2\}$  and ${J_i}({u^1}) < {J_i}({u^2})$, for some $ i \in \{ 1,2\}$.}


{\textbf{Problem 1.} Consider the nonlinear system (\ref{eq:1}) associated with the cost functions (\ref{eq:2}). Design the control policy $u(t)$, so that not only the states of the closed-loop system converge to the origin but also the cost functions (\ref{eq:2}) are minimized in Pareto sense. That is,
\begin{align}
\begin{array}{l}
\mathop {\min }\limits_u \,{\cal V}(J\left( {x,u} \right))\\
s.t.\,\,\,\,\,\,\dot x = f(x) + g(x)u
\end{array} \label{eq:04}
\end{align}
where  $J(.): = {[{J_1}(x,u),{J_2}(x,u)]^T}$ and ${\cal V}(.) \in \Re$  is a continuous value function.



\begin{figure}
\begin{center}
\includegraphics[width=6.5 cm]{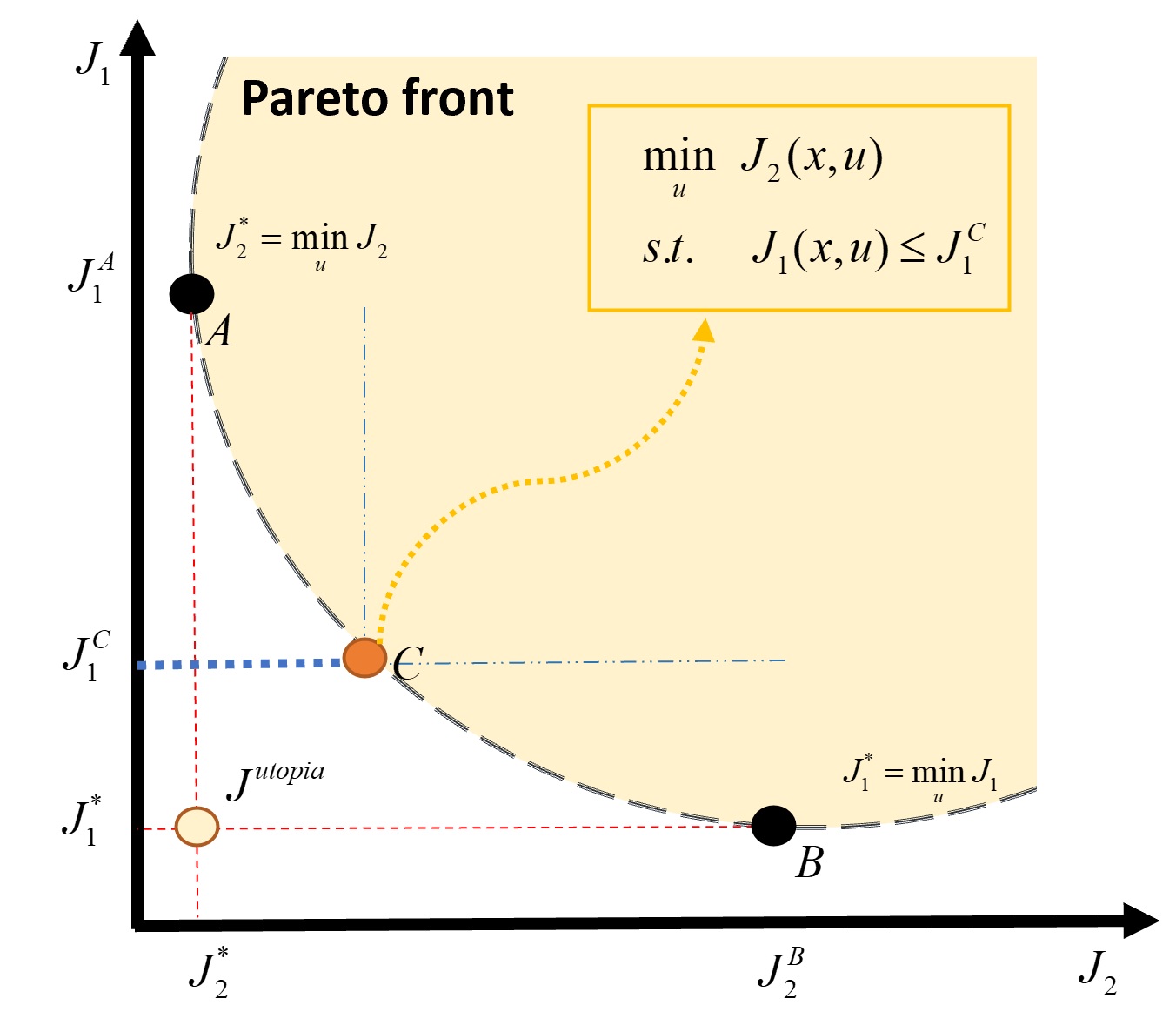}  
\caption{Schematic representation of Pareto optimal solutions for a bi-objective problem where $\{{\cal Y}\}_1$ is chosen as $J_1^C$.} \label{Fig98}
\end{center}
\end{figure}

 {\textbf{Definition 4}. For the system (\ref{eq:1}) with two objectives given by (\ref{eq:2}), a  control policy ${u^*} \in \Phi $ is called efficient (i.e., optimal in Pareto sense) if and only if $J\left( {x,{u^*}} \right) \le J\left( {x,u} \right)$, $\forall u \in \Phi $.}

{Problem 1 is completely equivalent to
\begin{align}
\begin{array}{l}
\mathop {\min }\limits_{u,{\cal Y} \in {\cal F}} \,{\cal V}({\cal Y})\\
s.t.\,\,\,\,\,\,\dot x = f(x) + g(x)u
\end{array} \label{eq:06}
\end{align}
where ${\cal F} \subset {\Re ^2}$  is the criteria space and defined as follows
\begin{align}
{\cal F}: = \{ {\cal Y} \in {\Re ^2}:{\cal Y} = J(x,u) \ge {J^{utopia}},\,\forall u \in \Phi \}.  \label{eq:07}
\end{align}}

{\textbf{Definition 5}. If $\exists u \in \Phi$, such that $J(x,u) \le {\cal Y}$, then  ${\cal Y} \in {\cal F}$ is called satisficing.}


{Let 
\begin{align}
{\cal C}: = \{ {\cal Y} \in {\Re ^2}:J(x,u) \le {\cal Y},\,for\,some\,\,u \in \Phi \}  \label{eq:08}
\end{align}
be the whole set of satisficing solutions. Note that ${\cal C}$ is the projection of $J(x,u) \le {\cal Y}$  with  $u \in \Phi $ onto the space of the ${\cal F}$.} 


{\textbf{Proposition 1.} Provided that ${{\cal Y}^*}$ is the optimal solution of
\begin{align}
\begin{array}{l}
\mathop {\min }\limits_{u, {\cal Y} \in {\cal F} \cap {\cal C}} \,{\cal V}({\cal Y})\\
s.t.\,\,\,\,\,\,\dot x = f(x) + g(x)u,
\end{array} \label{eq:09}
\end{align}
then ${{\cal Y}^*}$ is an optimal solution of Problem 1 in Pareto sense.}

{\textbf{Proof.} Assume that ${{\cal Y}^*}\in {\cal F} \cap {\cal C}$ is the solution of (\ref{eq:09}). This implies that  $\exists {u^*} \in \Phi $ such that ${{\cal Y}^*} = J(x,{u^*})$. Note that since ${{\cal Y}^0} \in {\cal F} \cap {\cal C}$  and ${\cal V}({{\cal Y}^0}) < {\cal V}({{\cal Y}^*})$, ${{\cal Y}^0} = J(x,{u^*})$  is contradictory to the optimality of  ${{\cal Y}^*}$, if  ${{\cal Y}^*} > J(x,{u^*})$.      Therefore, one has ${{\cal Y}^*} = {{\cal Y}^0} = J(x,{u^*})$ . Note also that $J(x,{u^*}) \in {\cal F}$  ( since  ${\cal F} \subset {\cal Y} \cap {\cal C}$) which implies that (\ref{eq:09}) and (\ref{eq:06}) are equivalent. This completes the proof. \hfill $\square$}

As an special case,  (\ref{eq:09}) can be rewritten as 
\begin{align}
&\mathop {\min }\limits_u \,{J_1}(x,u)  \nonumber \\
&s.t.\,\,\dot x = f(x) + g(x)u \nonumber \\
&\,\,\,\,\,\,{J_2}(x,u) \le {{\{\cal Y\}}_2} \label{eq:010}
\end{align}
where ${{\{\cal Y\}}_2}$ is an aspiration level on the cost functional $J_2$. Problem (\ref{eq:010}) is an extension of static $\varepsilon $-constraints optimization to dynamics optimizations.  As shown in Proposition 1, solution of $\varepsilon$-constraints problem (\ref{eq:010}) is on the Pareto front. (See {Fig.~\ref{Fig98}) 



Define the value function for a control policy $u \in \Phi $ as
\begin{align}
\Scale[1]{{V_i}(x(t)) = \int\limits_t^\infty  {{r_i}(x(\tau ),\;u)d\tau } ,\,\,i = 1,2,} \label{eq:3}
\end{align}
where ${V_i}(x(\infty )) = 0$.

Next, for an associated admissible policy $u \in \Phi $, define the Hamiltonian functionals  corresponding to the value functions (\ref{eq:3}) as
\begin{align}
\Scale[0.9]{ {{\cal H}_i}\;(x,u,{V_i}) := {Q_i}(x) + {u^T}{R_i}u + \nabla V_i^T\;(f(x) + g(x)u),}\label{eq:4}
\end{align}
for $i = 1,2$, where $\nabla {V_i}$ is the gradient of ${V_i}$. Note that the Bellman equation (\ref{eq:4}) is a differential equivalent to the value function (\ref{eq:3}).

The Pareto optimal vector ${\cal H}^*$ \cite{Lopez:2019} is defined as
\begin{align}
{{\cal H}^*}\;(x,u^*,{V}) :=& \mathop {\min }\limits_{u \in \Phi } {{\cal H}}\;(x,\;u,{V}) \label{eq:100}  \\ 
& { \,\,s.t. \,\,\dot x = f(x) + g(x)u} \nonumber  
\end{align}
or equivalently (by reformulating as an $\varepsilon $-constraints) as 
\begin{align}
{{\cal H}^*}\;(x,u^*,{V}) =&  \mathop {\min }\limits_{u \in \Phi } {\cal H}_1^{}(x,u(.),V_1) \quad \quad \quad \label{eq:08.a} \\ 
&{ \,\,s.t. \,\,\dot x = f(x) + g(x)u} \nonumber \\ 
& \quad -\delta(x) \le {\cal H}_2^{}(x,u(.),V_2) \le 0   \label{eq:08.b}
\end{align}
where ${\cal H}: = {[ {\cal H}_1, {\cal H}_2]^T}$, $V:= {[ V_1, V_2]^T}$ and $\delta(x)$ is the aspiration (i.e., the level of the performance) for objective 2 such that $\delta(x)  \le  \delta$, $\forall x \in \Omega  \subset {\Re ^n}$, $\delta > 0$, and $\Omega$ is the closed compact set containing the origin.



\textbf{Remark 1.}  {If there exists $u^*$ such that ${\cal H}^*(x)=[0,0]^T$  $ \forall x$, then $u^*$  (called utopia solution here and is introduced in \cite{Lopez:2019}) minimizes all value functions simultaneously and dominates all other admissible control policies. This implies that ${\cal H}^*=[0,0]^T$ is equivalent to ${J^{utopia}}$ (See Fig.~\ref{Fig98}). Moreover, note that under existence of a utopia solution, ${{\cal H}^*}$ is Pareto optimal in the sense that  ${{\cal H}^*}(x,{u^*},V) \le {\cal H}(x,u,V)$, $\forall u \in \Phi $,  $ \forall x$ and $\forall V = {[{V_1},{V_2}]^T}$  \cite{Lopez:2019}. However, for dynamic MO optimal control problems, normally ${{\cal H}^*}=[0,0]^T$ is unattainable, as conflict between objective arises over time. Note also that ${{\cal H}}=[0,0]^T$ occurs only for $u^*$ and for the pair $V_1^*$ and $V_2^*$ (See \cite{Lopez:2019} for more details). }

\textbf{Remark 2.} It is also worth noting that a smaller the aspiration $\delta(x)$ in (\ref{eq:08.b}) implies a better performance for $V_2(x)$ and imposing a looser aspiration on $V_2(x)$ means that we care less about it. In general case, if we have several objective functions, the values of aspiration levels show the priorities on their corresponding objective functions. Note also that the aspirations can be selected either based on some prior knowledge on the level of satisfaction of objectives in different scenarios or by a higher layer optimization framework that selects these aspirations to resolve the conflict between objectives or fix aspirations for some objectives and allow to relax the aspiration for some less-important ones while minimizing the relaxation factor.

\section{A Hamiltonian-driven Satisficing MO optimal control framework}

In this section, it is shown that the Hamiltonian functional corresponding to each objective can serve as a comparison function to compare the performance of admissible policies in a Pareto sense. The following theorem shows that minimizing one objective function while converting the other objective as a constraint resembles the satisficing (good enough) decision making framework for which the constraint bound is an indication of the aspiration level (the level of satisfaction) of the other objective function. 

\begin{thm}\label{theorem:1}
({Comparison theorem}) Let ${u^j}(.)$, $j = 1,2$ be two different admissible policies, with their value function vectors given as $\Scale[0.9]{{V^j}(x) = {[ {\begin{array}{*{20}{c}}
{V_1^j(.)}&{V_2^j(.)}
\end{array}} ]^T}}$, $j = 1,2$, where $V_i^j(.), i=1,2$  being the solution of ${{\cal H}_i(u^j,V_i^j)}=0, i=1,2$. Consider the following aspiration-satisfying dynamic optimization problem
\begin{align}
{\bar u^j}: =& \arg \min {\cal H}_1^{}(x,u(.),\;V_1^j) \quad \quad \quad \label{eq:8.a} \\
& { \,\,s.t. \,\,\dot x = f(x) + g(x)u} \nonumber \\ 
&- \delta^j(x) \le {\cal H}_2^{}(x,u(.),\;V_2^j) \le 0 \label{eq:8.b}
\end{align}
with $\delta^j(x)$ is the aspiration for objective 2 such that $\delta^j(x)  \le  \delta^j$, $\forall x \in \Omega  \subset {\Re ^n}$, $\delta^j > 0$, and $\Omega$ is the closed compact set including the origin. Let also ${\cal H}_{\min }^j: = {\left[ {\begin{array}{*{20}{c}}
{{\cal H}_1^j}&{{\cal H}_2^j}
\end{array}} \right]^T}$ where ${\cal H}_1^j: = {{\cal H}_1}(x,\bar u_{}^j(x),\;V_1^j)$ and ${\cal H}_2^j: = {{\cal H}_2}(x,\bar u_{}^j(x),\;V_2^j)$. Then, the following properties hold, $\forall x \in {\Omega}$:\\
1) ${\cal H}_{\min }^j \le {0_2}$, $j = 1,2$;\\
2)	If $- \,\delta^j(x) \le {\cal H}_2^j \le 0_2, j=1,2$, and ${\cal H}_1^1 < {\cal H}_1^2$, then $V_1^2 < V_1^1$ and consequently  ${V^1}{ \nprec }{V^2}$;\\
3)	For sufficiently small $\delta^1$ and $\delta^2$, if $\delta^2 < \delta^1$ and ${\cal H}_1^1 < {\cal H}_1^2$, then ${V^1}{ \nprec }{V^2}$  and  ${V^2}{ \nprec }{V^1}$.
\end{thm}
  
\textbf{Proof}. 
The proof has three parts. { It follows from (\ref{eq:8.b}) that ${\cal H}_2^j = {{\cal H}_2}(x,\bar u_{}^j(x),\;V_2^j) \le 0$. Moreover, based on ${\cal H}_1^{}(x,{u^j}(.),\:V_1^j) = 0$ and (\ref{eq:8.a}) which implies that ${\cal H}_1^j = \mathop {\min }\limits_u {\cal H}_1^{}(x,u,\:V_1^j)$, one has ${\cal H}_1^j \le {\cal H}_1^{}(x,{u^j}(.),\:V_1^j) = 0$. This proves part 1. } 
 We now prove part 2. {Let $V_1^2(x) = V_1^1(x) + {\Lambda ^1}$ where ${\Lambda ^1}:={\Lambda ^1}(x,{u^1(x)})$. Based on the Hamiltonian (\ref{eq:4}) for $V_1^1(.)$ and the stationary condition \cite{Lewis:2012}, one has}
\begin{align}
 &\Scale[0.95]{{\cal H}_1^2 = {Q_1}(x) + \nabla {V_1^2}^T(.)f(x) - \frac{1}{4}\nabla {V_1^2}^T{g^{}}(x)R_1^{ - 1}{g^T}(x)\nabla V_1^2} \nonumber \\
  &\,\,\,\,\,\,\,\,\, \Scale[0.9]{= \,\,{\cal H}_1^1\, + \frac{{d {\Lambda ^1} }}{{dt}}\, - \frac{1}{4}\nabla  {\Lambda ^1}^T{g^{}}(x)R_1^{ - 1}{g^T}(x)\nabla  {\Lambda ^1}}. \label{eq:5}
\end{align}
  After some manipulation, (\ref{eq:5}) can be rewritten as
\begin{align} 
  \Scale[1]{\frac{{d {\Lambda ^1} }}{{dt}} = {\cal H}_1^2 - {\cal H}_1^1 + \frac{1}{4}\nabla  {\Lambda ^1}^T{g^{}}(x)R_1^{ - 1}{g^T}(x)\nabla  {\Lambda ^1}}. \label{eq:6} 
\end{align} 
If ${\cal H}_1^2 - {\cal H}_1^1 \ge 0$, (\ref{eq:6}) implies that ${{d\ {\Lambda ^1} } \mathord{\left/
 {\vphantom {{d {\Lambda ^1}} {dt}}} \right.
 \kern-\nulldelimiterspace} {dt}} \ge 0$. Based on (3), $ {\Lambda ^1} (x(\infty)) = 0$, so (\ref{eq:6}) implies that $ {\Lambda ^1} \le 0$, $\forall x \in {\Omega}$. Thus, ${\cal H}_1^1 < {\cal H}_1^2$ implies that $V_1^2 < V_1^1$ and consequently ${V^1}{ \nprec }{V^2}$, $\forall x \in {\Omega}$. This completes the proof of part 2. To prove part 3, considering the inequality condition (\ref{eq:8.b}), the Lagrangian is  $\Scale[0.9]{{\Gamma ^j} = {\cal H}_1^{}(x,\bar u_{}^j(x),V_1^j(x)) + \lambda _{12}^j[ - {\cal H}_2^{}(x,\bar u_{}^j(x),V_2^j(x)) - \delta^j(x)]}$ where $\lambda _{12}^j$ is Lagrange multiplier. Provided that $\delta^1$  and $\delta^2$  are sufficiently small, from the Kuhn-Tucker condition \cite{Lewis:2012}, one can see that constraint (\ref{eq:8.b}) will be active, i.e., ${\cal H}_2^{}(x,\bar u_{}^j(x),V_2^j) = \delta^j$, $\lambda _{12}^j > 0$. Moreover, $\Scale[0.9]{\lambda _{12}^j =  - \partial H_1^{}(x,\bar u_{}^j(x),V_1^j(x))/\partial H_2^{}(x,\bar u_{}^j(x),V_2^j(x))}$ which based on property 2 indicates that an improvement in $\Scale[0.95]{{\cal H}_1^{}(x,\bar u_{}^j(x)\allowbreak,V_1^j(x))}$ may only be obtained at the cost of degradation in $\Scale[0.95]{{\cal H}_2^{}(x,\bar u_{}^j(x),V_2^j(x))}$. Therefore, the inequality condition (\ref{eq:8.b}) is active, i.e., $\Scale[0.95]{{\cal H}_2^j = {{\cal H}_2}(x,\bar u_{}^j(.),V_2^j) =} \linebreak  - \delta^j$ for $j = 1,2$. Thus, using property 2, $\delta^2 < \delta^1$ implies that  ${\cal H}_2^2 > {\cal H}_2^1$ which implies that  $V_2^1 < V_2^2$ and consequently  $\Scale[0.95]{{V^2}{ \nprec }{V^1}}$, $\forall x \in {\Omega}$. Moreover, from property 2, one has   $\Scale[0.95]{{\cal H}_1^1 < {\cal H}_1^2}$ implies that $\Scale[0.95]{V_1^2 < V_1^1}$ and consequently $\Scale[0.95]{{V^1}{ \nprec }{V^2}}$. This completes the proof.\hfill $\square$

\textbf{Assumption 1}. {Consider the nonlinear system (\ref{eq:1}). There exist feedback control policy $u(.)$ and functions ${V_{1}}(u(.)) \in {\cal P}$ and ${V_{2}}(u(.)) \in {\cal P}$, and $\delta (x) > 0$ such that} 
   \begin{align} 
  &0 \le {{\cal L}_1}({V_{1}}(.),u(.)) \nonumber \\
  &0 \le {{\cal L}_2}({V_{2}}(.),u(.)) \le \delta (x),\,\forall x \in {\Omega},   \label{eq:25}
   \end{align}
where, for any ${V_i} \in {C^1}$  and  $u \in \Phi $
 \begin{align} 
  {{\cal L}_i}({V_i},u) &=  - \nabla V_i^T(x)(f(x) + g(x)u) - {r_i}(x,u),\,i = 1,2 \nonumber \\ 
   &=-{{\cal H}_i}(x,u;V_i). \label{eq:26}  
  \end{align}

By Assumption 1 and Theorem 1, we immediately have the following corollary.

{\textbf{Corollary 1.} Consider the optimization problem (\ref{eq:08.a})-(\ref{eq:08.b}). Let Assumption 1 hold. Then, the following properties hold:\\ 
1) The aspiration-satisfying dynamic optimization problem (\ref{eq:08.a})-(\ref{eq:08.b}) is equivalent to $\varepsilon$-constraints problem (\ref{eq:010}); \\
2)  The value functions $V_1$ and $V_2$ are bounded.}

{ \textbf{Proof}.  1) By definition and based on (\ref{eq:3}), $V(x)=J(x,u)$ is the value function corresponding to the control policy $u$ and thus optimizing the cost function in (\ref{eq:04}), i.e., $\min _{u} J_1(x(0),u)$  is equivalent to $\min _{u} V_1(x)$. On the other hand, based on Theorem 1, this optimization amounts to (\ref{eq:08.a}), which shows that the objective function in (\ref{eq:04}) is equivalent to the objective function in (\ref{eq:08.a}). On the other hand, Theorem 1 shows that both optimization problem (\ref{eq:08.a})-(\ref{eq:08.b}) and (\ref{eq:010}) have the same feasible set of the Pareto front solutions. Therefore, both optimization problems are equivalent.  2) Based on (\ref{eq:4}), Property 1) of Theorem 1 and Assumption 1 imply that $V_1$ and $V_2$ are well-defined Lyapunov functions for the closed-loop system (\ref{eq:1}) and therefore the closed-loop system (\ref{eq:1}) is asymptotically stable. This implies that the value functions (i.e., $V_1$ and $V_2$), and accordingly, objective functions $J_1$ and $J_2$ are bounded. This completes the proof. \hfill $\square$ }

\textbf{Remark 3}.  {Fig.~\ref{Fig 11} provides us with an intuition that one can compare  different admissible policies by using their corresponding Hamiltonians as a measure. Based on Theorem 1, since ${\cal H}_{\min }^1({u_1})$ dominates ${\cal H}_{\min }^2({u_2})$  in Pareto sense, i.e., ${\cal H}_{\min }^2({u_2}) \prec {\cal H}_{\min }^1({u_1})$, ${V^1}: = {\rm{ }}{[V_1^1,V_2^1]^T}$ dominates ${V^2}: = {[V_1^2,V_2^2]^T}$ in Pareto sense where $V_1^1: = V_1^{}({u_1})$, $V_1^2: = V_1^{}({u_2})$, $V_2^1: = V_2^{}({u_1})$, and $V_2^2: = V_2^{}({u_2})$. That is ${V^2} \succ {V^1}$. Consequently, it implies that the admissible policy ${u^1}$ provides a better solution in terms of performance.} However, since ${\cal H}_{\min }^3({u_3})$ is not dominate ${\cal H}_{\min }^1({u_1})$ and ${\cal H}_{\min }^2({u_2})$ in Pareto sense, and vice versa, admissible policies ${u^k}$, $k = 1,2$ and ${u^3}$ are indifferent to each other.
\begin{figure}
\begin{center}
\includegraphics[width=7.5 cm]{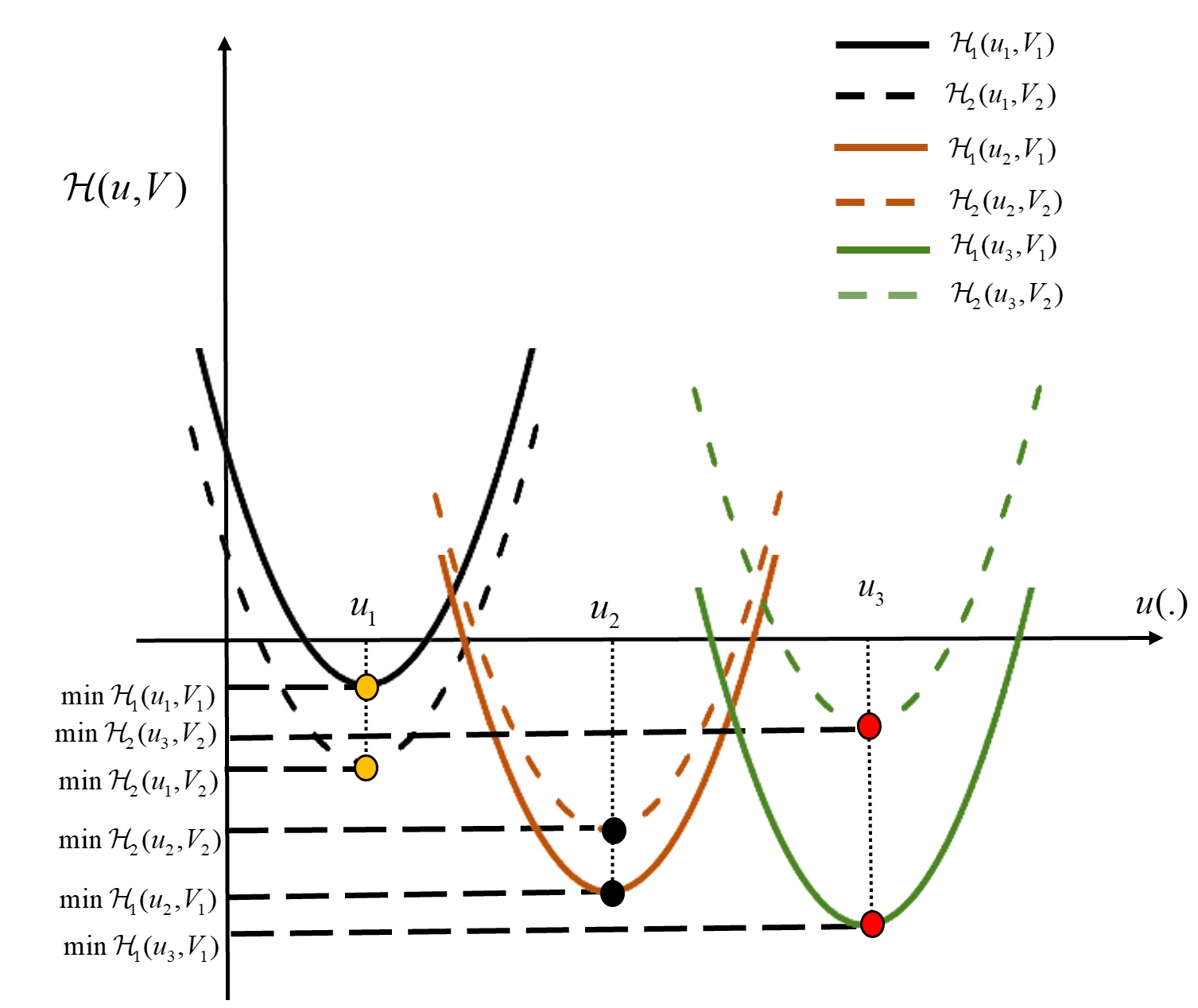}  
\caption{Comparison between three different admissible policies based on their Hamiltonian values.} \label{Fig 11}
\end{center}
\end{figure}

\textbf{Remark 4}. Based on property 3) of Theorem 1, different aspiration levels are corresponding to different Pareto optimal solutions.  Therefore, by tightening or loosing the aspiration level, i.e., $\delta^j(x)$, one can find different Pareto optimal solutions on the Pareto front, each corresponding to different demands on the second objective function.  The desired aspiration level might depend on the circumstance the system is encountering. Using this sense, in the next section, the problem in hand is formulated as an aspiration-satisfying optimization problem with HJB inequality as constraints.


\section{MULTI-OBJECTIVE CONTROL USING HJB INEQUALITIES}

In this section, Problem 1 is formulated as an aspiration-satisfying optimization problem with HJB inequalities as constraints. To this end, MO optimal control Problem 1 can be reformulated as the following aspiration-satisfying optimization problem.

\textbf{Problem 2}. Consider the nonlinear system (\ref{eq:1}) associated with the cost functions (\ref{eq:2}). Design the control policy $u(x)$, to solve the following aspiration-satisfying problem (\ref{eq:24.a})-(\ref{eq:24.d}).
\begin{align}  
&\Scale[1]{\mathop {\min }\limits_{ {u}} {\mkern 1mu} {\mkern 1mu} {\mkern 1mu} \int_\Omega  {{V_1}(x)dx}} \label{eq:24.a} \\
& \,\,s.t. \,\,\dot x = f(x) + g(x)u \nonumber \\ 
& \qquad\Scale[.95] {{{\cal H}_1}(x,u(V_1),V_1^{}) \le 0} \label{eq:24.b} \\
&\qquad \Scale[0.95]{ - \delta (x) \le {{\cal H}_2}(x,u(V_1),\;V_2(u)) \le 0} \label{eq:24.c} \\
&\qquad \Scale[0.95]{{V_i}(.)\, \in {\cal P},\,i = 1,2 } \label{eq:24.d}
\end{align}
where $\delta (x) > 0$ implicitly indicates the aspiration on optimizing objective ${V_2}$. Moreover, $\Omega  \subset {\Re ^n}$ is an arbitrary closed compact set including the origin that describes the region in which the objective function ${V_1}(x)$ is expected to be minimized the most.


\textbf{Remark 5}. Note that to solve the optimization problem $\mathop {\min }\limits_{u \in \Phi } {\cal H}_1^{}(x,u(.),V_1)$, one can reformulate it as the following equivalent problem
\begin{align}  
&\Scale[1]{\mathop {\min }\limits_{u \in \Phi } {\mkern 1mu} {\mkern 1mu} {\mkern 1mu} \int_\Omega  {{V_1}(x)dx}} \label{eq:024.a} \\
&\,\,s.t. \,\,\dot x = f(x) + g(x)u \nonumber \\ 
& \qquad  \Scale[.95] {{{\cal H}_1}(x,u(V_1),V_1^{}) \le 0} \label{eq:024.b} \\
&\qquad \Scale[0.95]{{V_1}(.)\, \in {\cal P} }. \label{eq:024.d}
\end{align}
 Moreover, based on Theorem 1, (\ref{eq:24.a}) -(\ref{eq:24.c}) are equivalent to (\ref{eq:8.a})-(\ref{eq:8.b}) which indicates that the cost functions (\ref{eq:2}) are minimized in a Pareto sense.





\begin{thm}\label{theorem:2}
  Let ${V_{1}} \in {\cal P}$  and its corresponding control policy ${u}$ be the solution of ${{\cal H}_1({u},{V_{1}})}=0$. Let Assumption 1 hold for the cost function ${V_{1}}({u}(.)) \in {\cal P}$  and ${V_{2}}({u}(.)) \in {\cal P}$, and control policy ${u}$. For a fixed $\delta (x) > 0$, the following hold.\\
1)	The aspiration-satisfying optimization Problem 2 has a nonempty feasible set.\\
2)	Let ${\bar V_1}({\bar u}(.)) \in {\cal P}$ and ${\bar V_2}({\bar u}(.)) \in {\cal P}$  be a feasible solution of the constrained optimization Problem 2. Then, the control policy ${\bar u}(.)$ is stabilizing. \\
3) For sufficiently large $\delta (x) > 0$, ${V_{1}}({u}(.))$, and ${V_{2}}({u}(.))$ is a optimal solution of the constrained optimization Problem 2.
\end{thm}

\textbf{Proof}.
The proof has two parts.\\
1)	Under Assumption 1 and Theorem 1, it is straightforward that the feasible set is not empty.\\
2)	For a feasible solution ${\bar V_1}({\bar u}(.)) \in {\cal P}$ and ${\bar V_2}({\bar u}(.)) \in {\cal P}$, the inequality equations (\ref{eq:24.b})-(\ref{eq:24.c}) are satisfied. It follows from (\ref{eq:26}) that
 \begin{align} 
{\dot {\bar{V_i}}} &= \nabla {\bar V_i}^T({\bar u}(.))(f(x) + g(x){\bar u}(.)) \nonumber \\
 &=  - {{\cal L}_i}({\bar V_i},{\bar u}(.)) - {r_i}(x,{\bar u}(.))\label{eq:28.a}
  \end{align}
which implies that if ${{\cal L}_1}({\bar V_1},{\bar u}(.)) \ge 0$ and ${{\cal L}_2}({\bar V_2},{\bar u}(.)) \ge 0$, then ${\bar V_1}$ and ${\bar V_2}$ are well-defined Lyapunov functions for the closed-loop system (\ref{eq:1}).   \\
3)	It follows from (\ref{eq:24.a})-(\ref{eq:24.c}) that for sufficiently large $\delta (x) > 0$, ${V_{1}} \in {\cal P}$ and its corresponding control policy ${u(V_{1})}$ are the solution of the constrained optimization Problem 2. Now, it is remaining to show that this solution is the optimal solution of optimization Problem 2.
Using (\ref{eq:1}) and (\ref{eq:28.a}), one has
\begin{align}
{{\dot {\bar V}}_1} &= - {{\cal L}_1}({{\bar V}_1},{{\bar u}}(.)) - {r_1}(x,{{\bar u}}(.)) \nonumber \\
{{\bar V}_1}({x_0}) &= \int\limits_0^T {{{\cal L}_1}({{\bar V}_1},{{\bar u}}(.)) + {r_1}(x,{{\bar u}}(.))]} dt + {{\bar V}_1}(x(T)) \nonumber \\
 {{\bar V}_1}({x_0}) &\ge \int\limits_0^T {{r_1}(x,{{\bar u}}(.))} dt + {{\bar V}_1}(x(T))
\end{align}
 Now, let $T \to  + \infty $. It follows from property 2 that ${\bar V_1}({x_0}) \ge {J_1}({x_0},{\bar u_1}(.))$.  Therefore, one has
\begin{align}
{\bar V_1}({x_0}) \ge {J_1}({x_0},{\bar u}(.)) \ge \mathop {\min }\limits_u {J_1}({x_0},u(.)) \mathop  = \limits^{\,\,\,}  {V_{1}}({x_0})
\end{align}
which implies that for any other feasible solution of Problem 2, i.e., ${{\bar V}_1}(x)$, one has ${{\bar V}_1}(x) \ge {V_{1}}(x)$ and consequently $\int\limits_\Omega  {{{\bar V}_1}(x)dx}  \ge \int\limits_\Omega  {{V_{1}}(x)dx}$. This completes the proof. \hfill $\square$

\section{SOS BASED MULTI-OBJECTIVE CONTROL}

In this section, a novel iterative method is developed to find the solution of Problem 2 and accordingly Problems 1 based on the Sum-of-Squares (SOS) based methods \cite{Ahmadi:2018}. To do so, the following definition is needed.

\textbf{Definition 6}. A polynomial $p(x)$ is an SOS polynomial, i.e., $p(x) \in {{\cal P}^{SOS}}$ where ${{\cal P}^{SOS}}$ is a set of SOS polynomial, if $p(x) = \sum\nolimits_1^m {p_i^2(x)} $ where $p_i^{}(x) \in {\cal P}$, $i = 1,...,m$. 

Let $\Scale[0.9]{{V_i}(x) = \sum\limits_{j = 1}^N {{c_{ij}}{m_{ij}}(x)}  = {C_i}^T\overrightarrow m_i^{(2,2d)}(x)}$, $i = 1,2$  where ${m_{ij}}(x)$, $i = 1,2$ are predefined monomials in $x$ and ${c_{ij}}$, $i = 1,2$ are coefficients to be determined. { Denote $V_i^k(x): = {C_i^k}^T\overrightarrow m_i^{(2,2d)}(x)$, $i=1,2$, $k=1,2,3,...$ where $k$ stands for iteration steps and $V_i^k(x)$,  $i = 1,2$ is the approximation of $V_i(x)$,  $i = 1,2$ at the iteration $k$-th.}

\textbf{Assumption 2.}  {For system (1), there exist polynomial functions ${V_1^0}$ and ${V_2^0}$, control policy ${u^0}(.)$, and aspiration level $\delta^0(x) \in {\cal P}^{SOS} $ such that  ${V_i^0} \in {\cal R}{\left[ x \right]_{2,2d}} \cap {{\cal P}^{SOS}}$, ${{\cal L}_i}({V_i^0}(.),u) \in {{\cal P}^{SOS}}$, and $\delta^0(x) - {{\cal L}_2}({V_2^0}(.),u) \linebreak \in {{\cal P}^{SOS}}$, $i = 1,2$.}

Motivated by the work in \cite{Jiang:2015} Algorithm 1 is given to find the solution of Problem 2.

\textbf{Remark 6}.  { Assumption 2 indicates that there exists a feasible solution for Algorithm 1.}


 \noindent\hrulefill\\
 {\bf Algorithm 1:} MO SOS based algorithm. \\
 \vspace{.02in}
 \noindent\hrulefill
 \vspace{.01in}
  \small
 	\begin{algorithmic}[1]
 	\Procedure{}{}
 		\State  {Initialize $\bar r = 0$ and $\delta^0(x)>0$.}
 		\State Set $k=1$ and start with $\{ V_1^0(.),V_2^0(.),u_{}^{(0)},\delta^{\bar r}(.)\} $ that satisfy Assumption 2.
 		\State If there is a feasible solution then solve the following SOS program:
 \begin{align}
 &\mathop {\min }\limits_{{K_{{c_1}}}} \,\,\,\,(\int\limits_\Omega  {\overrightarrow {m}_1^{(2,2d)}(x)dx{)^T}} {C_1} \label{eq:33.a}\\
 &s.t.\,\,\,\,\,\,\,{{\cal L}_i}(u(V_1),{V_i}(.))\, \in {{\cal P}^{SOS}},\,\,i = 1,2 \label{eq:33.b} \\
 & \,\,\,\,\,\,\,\,\,\,\,\,\delta^{\bar r}(x) - {{\cal L}_2}(u(V_1),{V_2}(u(V_1))\,\, \in {{\cal P}^{SOS}},\, \label{eq:33.c} \\
 &\,\,\,\,\,\,\,\,\,\,\,\,V_1^{k - 1} - V_1^{}\,\, \in {{\cal P}^{SOS}}, \label{eq:33.d} \\
 & \,\,\,\,\,\,\,\,\,\,\,V_i^{}\,\, \in {{\cal P}^{SOS}},i = 1,2,  \label{eq:33.f}
 \end{align} 
 where $V_i^{}(x): = C_i^T\overrightarrow{m}_i^{(2,2d)}(x)$, $V_i^k(x): = {C_i^k}^T\overrightarrow {m}_i^{(2,2d)}(x)$, $i = 1,2$, $u_{}^{}(V_1^{}) = K_{{C_1}}^{}\overrightarrow{m}_1^{(1,{{\bar d}^r})}$, $u_{}^{(k)}(V_1^{k}) = K_{{C_1}}^{k}\overrightarrow{m}_1^{(1,{{\bar d}^r})}$.

	\State  If $\left\| {C_1^k - C_1^{k - 1}} \right\|{ \leq }\gamma $, where $\gamma $ is a predefined threshold, or if there is no more feasible solution 
 		$u_r^* = u(V_1)$, ${U^*} = {U^*} \cup \left\{ {u_r^*} \right\}$ where  ${U^*}$ is the set of efficient control policies and go to Step 6 else go back to Step 4 with $k = k + 1$.
 		\State  {Set $\delta^{\bar r + 1}(x) = {\upsilon}\delta^{\bar r}(x)$,  where $0< {\upsilon} <1$ is predefined design parameter, then $\bar r = \bar r + 1$ and go to Step 3.	}
 	\EndProcedure
 	\vspace{.01in}	
 	\end{algorithmic}
 	\hrulefill
 \normalsize\\

\begin{thm}\label{theorem:3}
Assume that Assumptions 1-2 hold. Then, for a fixed aspiration level $\delta^{\bar r}(x)$, the following properties hold.\\
1)	The SOS program (\ref{eq:33.a})-(\ref{eq:33.f}) has at least one feasible solution;\\
2)	The control policy $u_{}^{(k + 1)}(x)$ is asymptotically stabilizing closed-loop system (\ref{eq:1}) at the origin;\\
3) $0 \le \,V_1^{k + 1} \le \,V_1^k$, $\forall k$, where $\,V_1^k\, \in {{\cal P}^{SOS}}$.\\
\end{thm}

\textbf{Proof}.
The proof has three parts. \\
 1)	Under Assumption 2 and Theorems 1-2, it is straightforward that SOS program (\ref{eq:33.a})-(\ref{eq:33.f}) has at least one feasible solution. \\
2) It follows from (\ref{eq:33.b}) that ${{\cal L}_1}(u(V_1),{V_1}(u^k(.)))\, \in {{\cal P}^{SOS}}$ and ${{\cal L}_2}(u(V_1),{V_2}(u^k(.)))\, \in {{\cal P}^{SOS}}$. 
Therefore, ${{\cal L}_1}(u(V_1),{V_1}(u^k(.)) \ge 0$ and ${{\cal L}_2}(u(V_1),{V_2}(u^k(.)) \ge 0$. It follows from (\ref{eq:26}) that
  \begin{align} 
{\dot {{V_i}}} &= \nabla { V_i}^T(u^k(.))(f(x) + g(x)u^k(.)) \nonumber \\
 &=  - {{\cal L}_i}({\bar V_i},u^k(.)) - {r_i}(x,u^k(.)) \nonumber \\
 & \le  - {Q_i}(x). \label{eq:23n}
  \end{align}
 {Since ${Q_i}(x)$, $i=1,2$, are positive definite, (\ref{eq:23n}) implies that ${V_1}$ and ${V_2}$ are well-defined Lyapunov functions for the closed-loop system.} Therefore, the control policy $u_{}^{(k)}(x)$ makes the system (\ref{eq:1}) asymptotically stabilizing at the origin.   \\
 3)	Constraints (\ref{eq:33.d}) and (\ref{eq:33.f}) imply $0 \le \,V_1^{k + 1} \le \,V_1^k$ and $\,V_1^k\, \in {{\cal P}^{SOS}}$, $\forall k$. This completes the proof.  \hfill $\square$



\textbf{Remark 7}. The proposed Algorithm 1 needs the complete knowledge of $f(x)$ and $g(x)$. However, the complete and precise system knowledge is not usually available in practical systems. Hence, in the next section, an online data-driven method is developed to employ the iterative scheme with real-time data.


\section{DATA-DRIVEN REINFORCEMENT LEARNING IMPLEMENTATION}

In this section, a data-driven satisficing reinforcement learning algorithm is developed to implement Algorithm 2 without having the full knowledge of the system dynamics. This algorithm uses measured data and a value function formulation with a span of a finite family of basis functions to approximate the solution of the problem 2.

Now, consider the system (\ref{eq:1}), after adding an exploratory probing noise, one has
\begin{align}\  
\dot x = f + g({u^{k}} + e) \label{eq:37}
 \end{align}
where  ${u^{k}}$ is a control policy at iteration $k$ and $e$ is an added bounded exploration probing noise.          

In the MO-SOS based Algorithm 1, under Assumption 2, one has $\forall k,r$, ${{\cal L}_i}(u_{}^k,V_i(.)) \in {\cal R}{\left[ x \right]_{2,2{{\bar d}^r}}}$, $i = 1,2$, $\delta^r(x) - {{\cal L}_2}(u_{}^k,V_2(.)) \in {\cal R}{\left[ x \right]_{2,2{{\bar d}^r}}}$, where $\delta^r(x) $, if the integer ${\bar d^r}$ satisfies
\begin{align}
\begin{array}{l}
{{\bar d}^r} \ge \frac{1}{2}\max \{ \deg (f(.)) + 2d - 1,2\deg (g(.))\\
\,\,\,\,\, + 2(2d - 1),\deg ({Q_1}(.))\, + \deg ({Q_2}(.)),\deg (\delta^r(.))\} 
\end{array} \label{eq:39}
 \end{align} 
where $\deg (.)$ represents the degree of the polynomial which is the highest degree of any of the terms. Also, ${u^{k}}$ obtained from the proposed MO-SOS based Algorithm 1 satisfies ${u^{k}} \in {\cal R}{\left[ x \right]_{1,{{\bar d}^r}}}$, $\forall k,r$.

Therefore, a matrix ${K_{{C_1}}^{k}} \in {\Re ^{m \times {n_{{{\bar d}^r}}}}}$, with $\Scale[0.9]{{n_{{{\bar d}^r}}} = \left( {\begin{array}{*{20}{c}}
{n + {{\bar d}^r}}\\
{{{\bar d}^r}}
\end{array}} \right) - 1}$ exists, such that ${u^{k}} = {K_{{C_1}}^{k}}\overrightarrow m_1^{(1,{{\bar d}^r})}$.
 Also, suppose there exist constant vectors ${C_1} \in {\Re ^{{n_{2d}}}}$  and  ${C_2} \in {\Re ^{{n_{2d}}}}$, with $\Scale[0.9]{{n_{2d}}{\rm{ }} = {\rm{ }}\left( {\begin{array}{*{20}{c}}
{n + 2d}\\
{2d}
\end{array}} \right) - n - 1}$, such that $\Scale[0.9]{V_1^{}(x): = C_1^T\overrightarrow m_1^{(2,2d)}(x)}$  and  $\Scale[0.9]{V_2^{}(x): = C_2^T\overrightarrow m_2^{(2,2d)}(x)}$. It follows then from (\ref{eq:37}) that
\begin{align}
&{{\dot V}_1} =  - {r_1}(x,{u^{k}}) - {{\cal L}_1}({u^{k}},V_1^{}(.))+ \Scale[0.9]{{(R_1^{ - 1}{g^T}\nabla {V_1})^T}{R_1}e} \label{eq:40.a} \\
&{{\dot V}_2} =  - {r_2}(x,{u^{k}}) - {{\cal L}_2}({u^{k}},V_2^{}(.)) \nonumber \\
& \quad \quad \quad \quad \quad \quad \quad \quad\quad \quad \Scale[0.9]{+ \nabla V_2^T\nabla V_1^{ - T}{(R_1^{ - 1}{g^T}\nabla {V_1})^T}R_1^{}e} \label{eq:40.b}
\end{align}

Notice that the terms ${{\cal L}_1}({u^{k}},V_1^{}(.))$, ${{\cal L}_2}({u^{k}},\linebreak V_2^{}(.))$, $R_1^{ - 1}{g^T}\nabla {V_1}$, and $\Scale[0.9]{\nabla V_2^T\nabla V_1^{ - T}{(R_1^{ - 1}{g^T}\nabla {V_1})^T}R_1 e}$  depend on the dynamic of the system. Also, note that constant vectors and matrix ${l_{{C_1}}} \in {\Re ^{{n_{2{{\bar d}^r}}}}}$  and ${l_{{C_2}}} \in {\Re ^{{n_{2{{\bar d}^r}}}}}$, and  $K_{{C_1}}^{} \in {\Re ^{m \times {n_{{{\bar d}^r}}}}}$ with $\Scale[0.9]{{\bar d^r} = \left( {\begin{array}{*{20}{c}}
{n + 2{{\bar d}^r}}\\
{2{{\bar d}^r}}
\end{array}} \right) - {\bar d^r} - 1}$  for the tuple $({V_1},{V_2},{u^{k}})$ can be chosen such that:
\begin{align}
&{{\cal L}_i}({u^{k}},V_i^{}(.)) = {l_{{C_i}}}^T\overrightarrow m_i^{(2,2{{\bar d}^r})}(x),\,\,i = 1,2,\label{eq:41.a} \\
&\Scale[0.95]{- \frac{1}{2}R_1^{ - 1}{g^T}\nabla {V_1} = K_{{C_1}}^{}\overrightarrow m_1^{(1,{{\bar d}^r})}}\label{eq:42.b}
\end{align}

Therefore, calculating ${{\cal L}_i}({u^{k}},V_i^{}(.))$, $i = 1,2$ and $R_1^{ - 1}{g^T}\nabla {V_1}$ amounts to find ${l_{{C_1}}}$, ${l_{{C_2}}}$, and $K_{{C_1}}^{}$.

Substituting (\ref{eq:41.a}) and (\ref{eq:42.b}) in (\ref{eq:40.a})-(\ref{eq:40.b}), we have
\begin{align}
&\Scale[0.95]{{\dot V_1} =  - {r_1}(x,{u^{k}}) - {l_{{C_1}}}^T\overrightarrow m_1^{(2,2{{\bar d}_r})}(x)- 2{(\overrightarrow m_1^{(1,{{\bar d}^r})})^T}K_{{C_1}}^T{R_1}e} \label{eq:43.a} \\
&\Scale[0.95]{{{\dot V}_2} =  - {r_2}(x,{u^{k}}) - {l_{{C_2}}}^T\overrightarrow m_2^{(2,2{{\bar d}_r})}(x)- 2(\nabla \overrightarrow m_2^{(2,2d)}(x(t)))_{}^T}  \nonumber \\
& \qquad \qquad  \Scale[0.95]{ \times C_2^{}{(\overrightarrow m_1^{(1,{{\bar d}^r})})^T}({(\nabla \overrightarrow m_1^{(2,2d)}(x(t)))^T}C_1^{})_{}^TK_{{C_1}}^T{R_1}e} \label{eq:43.b}
\end{align}

Integrating both sides of (\ref{eq:43.a})-(\ref{eq:43.b}) on the interval $\left[ {t,{\rm{ }}t + \delta t} \right]$ yields the following off-policy integral RL Bellman equations
\begin{align}
&\Scale[0.95]{C_1^T(\overrightarrow m_1^{(2,2d)}(x(t)) - \overrightarrow m_1^{(2,2d)}(x(t + \delta t))) =} \nonumber  \\
&\Scale[0.95]{\int\limits_t^{t + \delta t} {({r_1}(x,{u^{k}}) + {l_{{C_1}}}^T\overrightarrow m_1^{(2,2{{\bar d}_r})}(x) + 2{{(\overrightarrow m_1^{(1,{{\bar d}^r})})}^T}K_{{C_1}}^T{R_1}e)d\tau }} \label{eq:44.a} \\
&\Scale[0.95]{C_2^T(\overrightarrow m_2^{(2,2d)}(x(t)) - \overrightarrow m_2^{(2,2d)}(x(t + \delta t)))=} \nonumber \\
&\Scale[0.95]{ \int\limits_t^{t + \delta t} {({r_2}(x,{u^{k}}) + {l_{{C_2}}}^T \overrightarrow m_2^{(2,2{{\bar d}_r})}(x)}+ 2(\nabla \overrightarrow m_2^{(2,2d)}(x(t)))_{}^T}  \nonumber  \\
&\quad \Scale[0.95]{\times C_2^{}{(\overrightarrow m_1^{(1,{{\bar d}^r})})^T}({(\nabla \overrightarrow m_1^{(2,2d)}(x(t)))^T}C_1^{})_{}^TK_{{C_1}}^T{R_1}e)d\tau} \label{eq:44.b}
\end{align} 

It follows from (\ref{eq:44.a})-(\ref{eq:44.b}) that ${l_{{C_1}}}$, ${l_{{C_2}}}$, and $K_{{C_1}}^{}$ can be found by using only the information of the system trajectories measured during a time interval, without requiring any system dynamic information. To this end, we define the following matrices:
\begin{align}
&\Scale[.95]{\sigma _e^1 = - {[ {\begin{array}{*{20}{c}}
{\overrightarrow m_1^{(2,2d)}}&{2{{(\overrightarrow m_1^{(1,{{\bar d}^r})})}^T} \otimes {e^T}{R_1}^{}}
\end{array}} ]^T}}, \label{eq:45.a}  \\
&\Scale[.95]{\begin{array}{l}
\sigma _e^2 = - [\begin{array}{*{20}{c}}
{\overrightarrow m_2^{(2,2d)}}&{2(\nabla \overrightarrow m_2^{(2,2d)}(x(t)))_{}^TC_2^{}{{(\overrightarrow m_1^{(1,{{\bar d}^r})})}^T} \times }
\end{array}\\
\qquad \quad \quad \qquad \quad \quad ({(\nabla \overrightarrow m_1^{(2,2d)}(x(t)))^T}C_1^{})_{}^T \otimes {e^T}{R_1}{]^T},
\end{array}} \label{eq:45.a.b} \\
&\Scale[.95]{\phi _i^{k} = {[ {\begin{array}{*{20}{c}}
{\int\limits_{{t_{0,k}}}^{{t_{1,k}}} {\sigma _e^id\tau } }&{}& \cdots &{\int\limits_{{t_{{q_{k}} - 1,k }}}^{{t_{{q_{k}},k}}} {\sigma _e^id\tau } }
\end{array}} ]^T}}, \label{eq:45.b} \\
&\Scale[.95]{\Xi _i^{k} = {[ {\begin{array}{*{20}{c}}
{\int\limits_{{t_{0,k}}}^{{t_{1,k}}} {{r_i}(x,{u^{k}})d\tau } }&{}& \cdots &{\int\limits_{{t_{{q_{k}} - 1,k }}}^{{t_{{q_{k}},k}}} {{r_i}(x,{u^{k}})d\tau } }
\end{array}} ]^T}} \label{eq:45.c} \\
&\Scale[.95]{\theta _i^{k} = {[ {\begin{array}{*{20}{c}}
{\overrightarrow m_i^{(2,2d)}\left| {_{{t_{0,k}}}^{{t_{1,k}}}} \right.}&{}& \cdots &{\overrightarrow m_i^{(2,2d)}\left| {_{{t_{{q_{k}} - 1,k}}}^{{t_{{q_{k}},k}}}} \right.}
\end{array}} ]^T},}\label{eq:45.d}
\end{align}
for $i = 1,2$, where $\phi _i^{k}\in {\Re ^{q_i^{k} \times ({n_{2{{\bar d}^r}}} + m{n_{{{\bar d}^r}}})}}$ and $\Xi _i^{k} \in {\Re ^{q_i^{k}}}$.

It follows from (\ref{eq:44.a})-(\ref{eq:44.b}) that 
\begin{align}
&\Scale[0.95]{\phi _1^{k}\left[ {\begin{array}{*{20}{c}}
{{l_{{C_1}}}}\\
{Vec(K_{{C_1}}^{})}
\end{array}} \right] = \Xi _1^{k} + \theta _1^{k}{C_1}} \label{eq:46.a} \\
&\Scale[0.95]{\phi _2^{k}\left[ {\begin{array}{*{20}{c}}
{{l_{{C_2}}}}\\
{Vec(K_{{C_1}}^{})}
\end{array}} \right] = \Xi _2^{k} + \theta _2^{k}{C_2}} \label{eq:46.b}
\end{align}

\textbf{Assumption 3}. At each iteration $k$, there exists a lower-bound $q_0^{k} \in \mathbb{Z}^ + $ such that if $q_1^{k},\,\,q_2^{k} \ge q_0^{k}$ where
 $q_1^{k}$ and $q_2^{k}$  are dimensional of vectors $\Xi _1^{k}$  and $\Xi _2^{k}$, respectively, then $rank(\phi _1^{k}) \linebreak = {n_{2{{\bar d}^r}}} + m{n_{{{\bar d}^r}}}$ and $rank(\phi _2^{k}) = {n_{2{{\bar d}^r}}} + m{n_{{{\bar d}^r}}}$.

Now, assume that $q_1^{k},\,\,q_2^{k} \ge q_0^{k}$, $\forall k$. It follows from (\ref{eq:46.a})-(\ref{eq:46.b}) that the values of ${l_{{C_1}}} \in {\Re ^{{n_{2{{\bar d}^r}}}}}$, ${l_{{C_2}}} \in {\Re ^{{n_{2{{\bar d}^r}}}}}$, and $K_{{C_1}}^{} \in {\Re ^{m \times {n_{{{\bar d}^r}}}}}$ are determined {using the least square method in average sense} as follows:
\begin{align}
\Scale[0.95]{\left\{ \begin{array}{l}
\left[ {\begin{array}{*{20}{c}}
{{l_{{C_1}}}}\\
{Vec(K_{{C_1}}^{})}
\end{array}} \right] = {({(\phi _1^{k})^T}\phi _1^{k})^{ - 1}}{(\phi _1^{k})^T}(\Xi _1^{k} + \theta _1^{k}{C_1})\\
\left[ {\begin{array}{*{20}{c}}
{{l_{{C_2}}}}\\
{Vec(K_{{C_1}}^{})}
\end{array}} \right] = {(\phi _2^{k})^T}\phi _2^{k}{)^{ - 1}}{(\phi _2^{k})^T}(\Xi _2^{k} + \theta _2^{k}{C_2})
\end{array} \right.} \label{eq:48}
\end{align}

\textbf{Remark 8}.  { Note that Assumption 3 implies that a necessary condition of persistency of excitation (i.e., rank conditions) for solving (\ref{eq:48}) is guaranteed.}

So, an iterative SOS based data-driven learning algorithm is proposed in Algorithm 2 for online implementation of Algorithm 1.


\noindent\hrulefill
\\
{\bf Algorithm 2:} Data-driven MO SOS based algorithm.
\\
\vspace{.02in}  
{\hrulefill
\vspace{.05in}
 \small
	\begin{algorithmic}[1]
	\Procedure{}{}
	\State {Initialize $\bar r = 0$ and $\delta^0(x)>0$.}
		\State Set $k=1$ and find the tuple $\{ V_1^0,V_2^0,u_{}^{(0)},\delta^{\bar r}\} $ such that Assumption 2 be satisfied. Choose $C_1^0$ and $C_2^0$ such that $V_1^0(x): = {(C_1^0)^T}\overrightarrow m_1^{(2,2d)}(x)$ and  $V_2^0(x): = {(C_2^0)^T}\overrightarrow m_2^{(2,2d)}(x)$.

		\State Employ $u = {u^k} + e$ as the input to the system (\ref{eq:1}), where $e$ is the probing noise and calculate and construct $\Xi _1^{},\,\,\Xi _2^{},\,\,\theta _1^{},$ and $\theta _2^{}$ as (\ref{eq:45.a})-(\ref{eq:45.d}), untill $\phi _1^{},\,\,\phi _2^{}$ be of full column rank.

		\State Solve the following SOS program to find an optimal solution $\{ C_1^k,{\rm{ }}C_2^k,K_{{C_1}}^k\}$:
		\begin{align}
		&\mathop {\min }\limits_{{K_{{c_1}}}} \,\,\,\,(\int\limits_\Omega  {\overrightarrow {m}_1^{(2,2d)}(x)dx{)^T}} {C_1} \label{eq:49.a}  \\
		&s.t. \Scale[0.85]{\left[ {\begin{array}{*{20}{c}}
				{{l_{{C_1}}}}\\
				{Vec(K_{{C_1}}^{})}
				\end{array}} \right] = {({(\phi _1^{k})^T}\phi _1^{k})^{ - 1}}{(\phi _1^{k})^T}(\Xi _1^{k} + \theta _1^{k}{C_1}),}\label{eq:49.b}  \\
		& \Scale[0.85]{\left[ {\begin{array}{*{20}{c}}
		{{l_{{C_2}}}}\\
		{Vec(K_{{C_1}}^{})}
		\end{array}} \right] = {(\phi _2^{k})^T}\phi _2^{k}{)^{ - 1}}{(\phi _2^{k})^T}(\Xi _2^{k} + \theta _2^{k}{C_2}),}\label{eq:49.c}  \\
		&\,\,\,\,\,\,\,\,\,\,\,{l_{{C_i}}}^T \overrightarrow m_i^{(2,2{{\bar d}_r})}(x) \in {{\cal P}^{SOS}},\,\,\,i = 1,2 \label{eq:49.d} \\
		&\,\,\,\,\,\,\,\,\,\,\,\delta^{\bar r}(x) - {l_{{C_2}}}^T \overrightarrow{m}_2^{(2,2{{\bar d}_r})}(x)\, \in {{\cal P}^{SOS}},\, \label{eq:49.f} \\
		&\,\,\,\,\,\,\,\,\,\,\,{(C_1^{k - 1} - C_1^{})^T} \overrightarrow{m}_1^{(2,2d)}(x)\,\, \in {{\cal P}^{SOS}}, \label{eq:49.g}
		\end{align}
		\State Update the value functions and control policy as follows:
		\begin{align}
		&V_i^k(x): = {C_i^k}^T \overrightarrow{m}_i^{(2,2d)}(x),\,i = 1,2 \label{eq:50.a} \\
		&u_{}^{(k)}(x) = K_{{C_1}}^{k}\overrightarrow{m}_1^{(1,{{\bar d}^r})} \label{eq:50.c}
		\end{align}
		\State If $\left\| {C_1^k - C_1^{k - 1}} \right\|{ \leq }\gamma $, where $\gamma $ is a predefined threshold, or if there is no more feasible solution  $u_r^* = u_{}^{(k)}(x)$ and go to Step 8 else go back to Step 4 with $k = k + 1$.
		\State  {Set $\delta^{\bar r + 1}(x) = {\upsilon}\delta^{\bar r}(x)$,  where $0< {\upsilon} <1$ is predefined design parameter, then $\bar r = \bar r + 1$ and go to Step 3.}
				
	\EndProcedure
	
	\vspace{.05in}
			\hrule
	\end{algorithmic}

\normalsize

\vspace{0.1in}

\begin{thm}\label{theorem:4}
Assume that Assumptions 1-3 hold. Then, for a fixed $\delta^{\bar r}(x)$, the following properties hold.\\
1)	There exists at least one feasible solution for the SOS program (\ref{eq:49.a})-(\ref{eq:49.g});\\
2)	The control policy $u_{}^{(k)}(x)$ (\ref{eq:50.c}) is asymptotically stabilizing the system (1) at the origin;\\
3)	$0 \le \,V_1^{k + 1} \le \,V_1^k$, $\forall k$, where $\,V_1^k\,$ is given in (\ref{eq:50.a}).
\end{thm}

\textbf{Proof}.
Provided that $\{ C_1^k,{\rm{ }}C_2^k\}$ is a feasible solution of the optimization problem (\ref{eq:33.a})-(\ref{eq:33.f}), one can find the corresponding matrix $K_{{C_1}}^k \in {\Re ^{m \times {n_{{{\bar d}^r}}}}}$  such that the tuple $\{ C_1^k,{\rm{ }}C_2^k,K_{{C_1}}^k\}$ be a feasible solution of the data-driven MO SOS program (\ref{eq:49.a})-(\ref{eq:49.g}) and (\ref{eq:50.a})-(\ref{eq:50.c}), which imply that property 1 holds. Moreover, since the tuple $\{ C_1^k,{\rm{ }}C_2^k,K_{{C_1}}^k\}$ is a feasible solution of the data-driven MO SOS program (\ref{eq:49.a})-(\ref{eq:49.g}) and (\ref{eq:50.a})-(\ref{eq:50.c}) and the tuple $\{ C_1^k,{\rm{ }}C_2^k\}$ is a feasible solution of the MO SOS program (\ref{eq:33.a})-(\ref{eq:33.f}) and Algorithms 1 and 2 have the equal objective function, $\Scale[0.9]{K_{{C_1}}^{k}\overrightarrow m_1^{(1,{{\bar d}^r})}}$ is an optimal solution of the MO SOS program (\ref{eq:33.a})-(\ref{eq:33.f}) and consequently the results of Theorem 3 are further extended to Theorem 4. This completes the proof. \hfill $\square$

\begin{figure}
\begin{center}
\includegraphics[width=8.2cm]{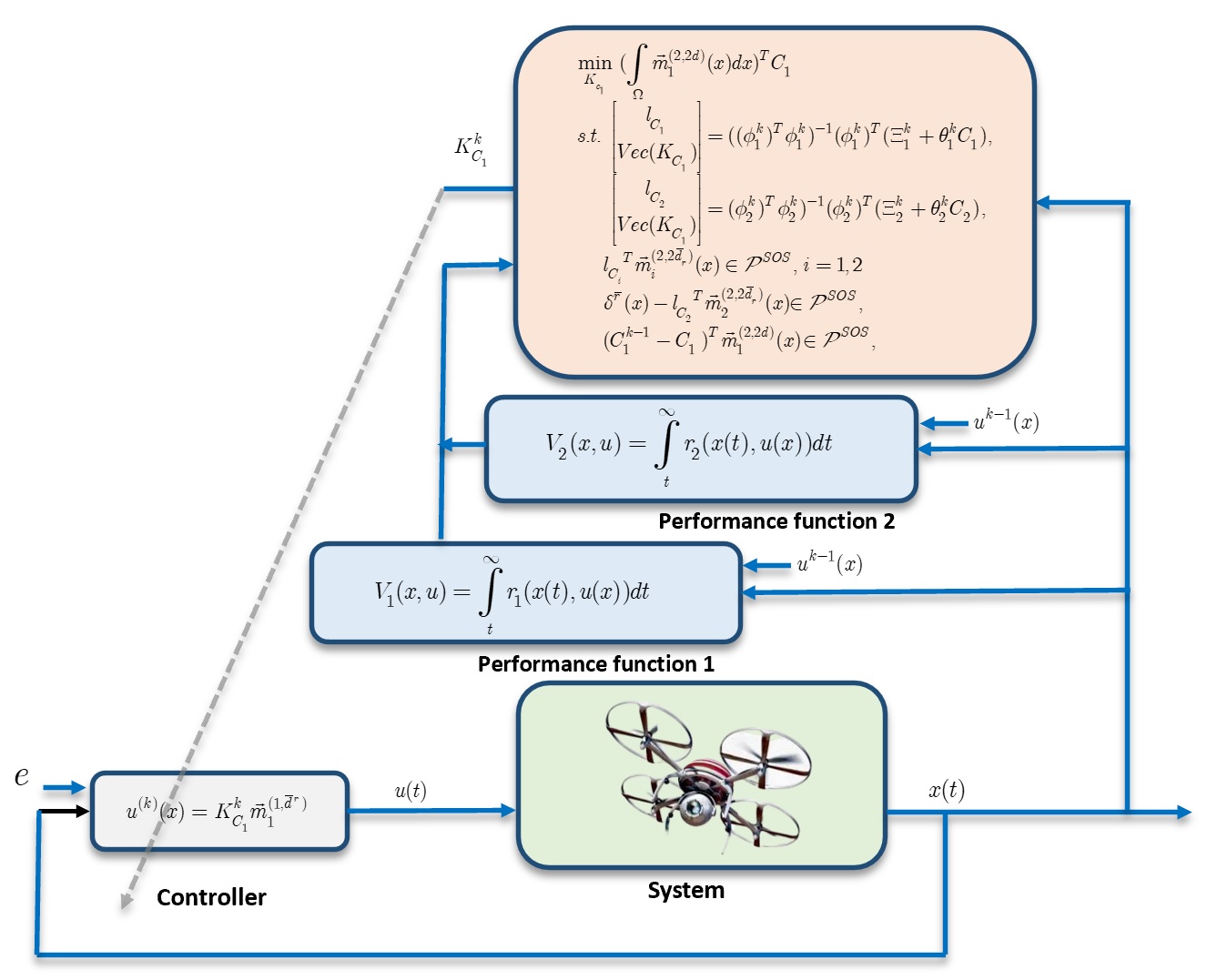}  
\caption{ {Proposed SOS-based data-driven learning control architecture.}} \label{Fig9}
\end{center}
\end{figure}

\section{SIMULATION}
In this section, the effectiveness of the proposed scheme is verified by two simulation examples.

\begin{figure}
\begin{center}
\includegraphics[width=6.2cm]{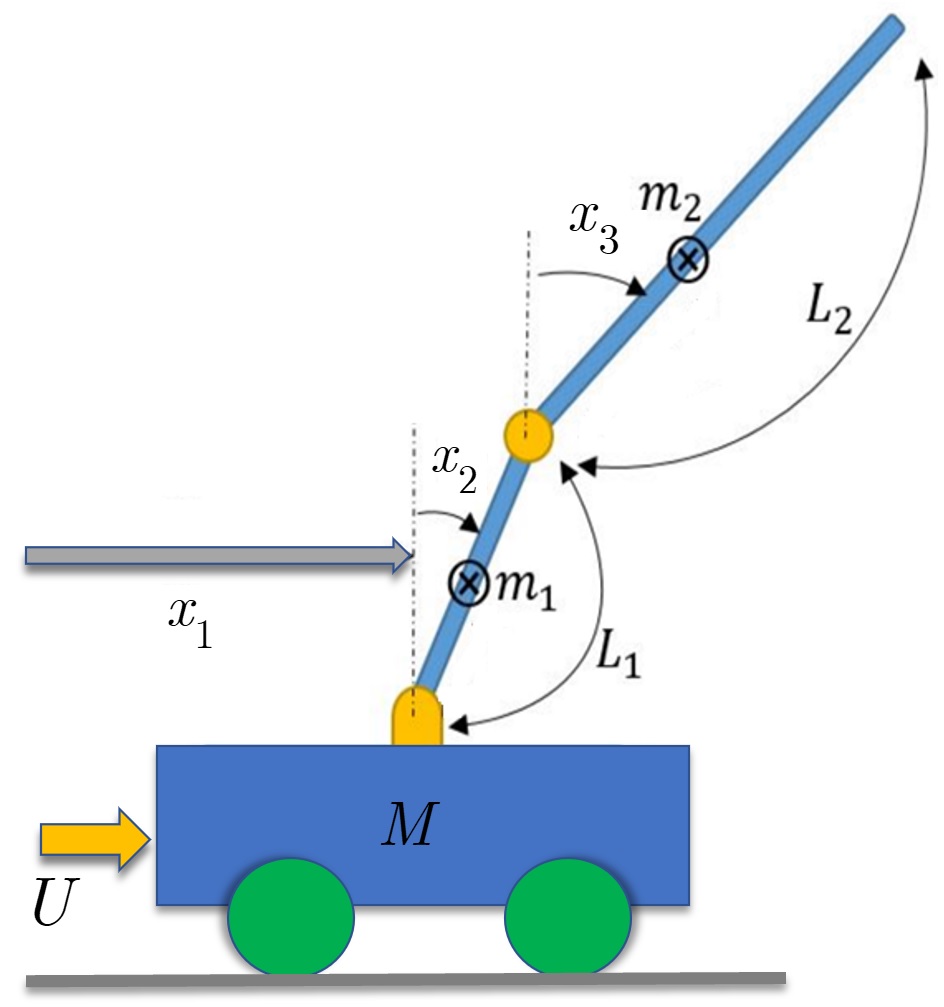}  
\caption{{Schematic of double inverted pendulum system. Specific values are: $l_1=0.18m$, $l_2=0.57m$, $M=0.43kg$, $m_1=0.08kg$, $m_2=0.11kg$, and $g = 9.86m/{s^2}$.}} \label{Fig0}
\end{center}
\end{figure}

\textbf{Example 1}. {Consider the linearized double inverted pendulum in a cart under the action of a single control input, shown in Fig.~\ref{Fig0}, with dynamics given by \cite{Lopez:2019} and \cite{Li:2008}}

\begin{align}
\dot X = \left[ {\begin{array}{*{20}{c}}
0&0&0&1&0&0\\
0&0&0&0&1&0\\
0&0&0&0&0&1\\
0&0&0&0&0&0\\
0&{86.69}&{ - 21.61}&0&0&0\\
0&{ - 40.31}&{39.45}&0&0&0
\end{array}} \right]X + \left[ {\begin{array}{*{20}{c}}
0\\
0\\
0\\
1\\
{6.64}\\
{0.08}
\end{array}} \right]U
\end{align}
{for the neighborhood of unstable balance point $X(t) = {[0,0,0,0,0,0]^T}$, where $X(t) = {[{x_1}(t),{x_2}(t),{x_3}(t),{\dot x_1}(t),{\dot x_2}(t),{\dot x_3}(t)]^T}$, with ${x_1}$, ${x_2}$, and ${x_3}$ are the position of the cart and angles of both pendulums, respectively; ${x_4}$, ${x_5}$, and ${x_6}$ are the velocities.} The quadratic cost functions chosen as  
\begin{align}
{J_i}({x_0},u) = \int_0^\infty  {({x^T}Q_ix + {u^T}R_i{u})} dt,i=1,2
\end{align}
with ${Q_1} = {I_6}$, ${Q_2} = 200*{I_6}$, and ${R_1} = {R_2} = 1$, with $\Omega  = \{ x\left| {x \in {\Re ^6} ,} \right.\,\left\| {{x}} \right\|_2 \le 1.7\}$. {Let us choose products of the set $\left\{ {{x_1},{\rm{ }}{x_2},{\rm{ }}{x_3},{\rm{ }}{x_4},{\rm{ }}{x_5},{\rm{ }}{x_6},{\rm{ }}x_1^2,{\rm{ }}x_2^2,{\rm{ }}x_3^2,{\rm{ }}x_4^2,{\rm{ }}x_5^2,{\rm{ }}x_6^2} \right\}$ with itself as the monomials for the value functions}. After the implementation of Algorithm 2 with three different aspiration levels as ${\delta^i} = \delta ^r(0.2x_1^2{x_3} + 0.1x_2^2{x_5} + 0.25x_4^2 + 0.2{x_2}{x_4}{x_6}  + 0.5{x_5}{x_6}+ 0.7x_1^2{x_4} + 0.2x_5^2 + 0.1{x_6}x_2^2 + 0.5{x_4}{x_5}{x_6} + 0.2{x_1}{x_2}{x_3})$, $i=1,2,3$ with $\delta^r \in \left\{ {0.001,0.14,2} \right\}$, three suboptimal control policies are obtained after three iterations. Fig.~\ref{Fig 1} shows the evolution of the system states after applying the obtained policies and the method in \cite{Lopez:2019}. It can be seen in Figs.~\ref{Fig 2}-\ref{Fig 1} that by changing the aspiration level on second objective the obtained control policies and corresponding system states are changed. That is, the trade-off between regulation error and control effort are changed by changing the aspiration level on second objective. Moreover, it can be seen in Fig.~\ref{Fig 1} that all the state trajectories are stabilized by the controller, which shows that the MO control problem 1 is solved in this case. Using Figs.~\ref{Fig 2}-\ref{Fig 1} one can compare the proposed method in this paper and the method in
\cite{Lopez:2019} regarding the system state trajectories and control policies. 

\begin{figure}
\begin{center}
\includegraphics[width=9 cm]{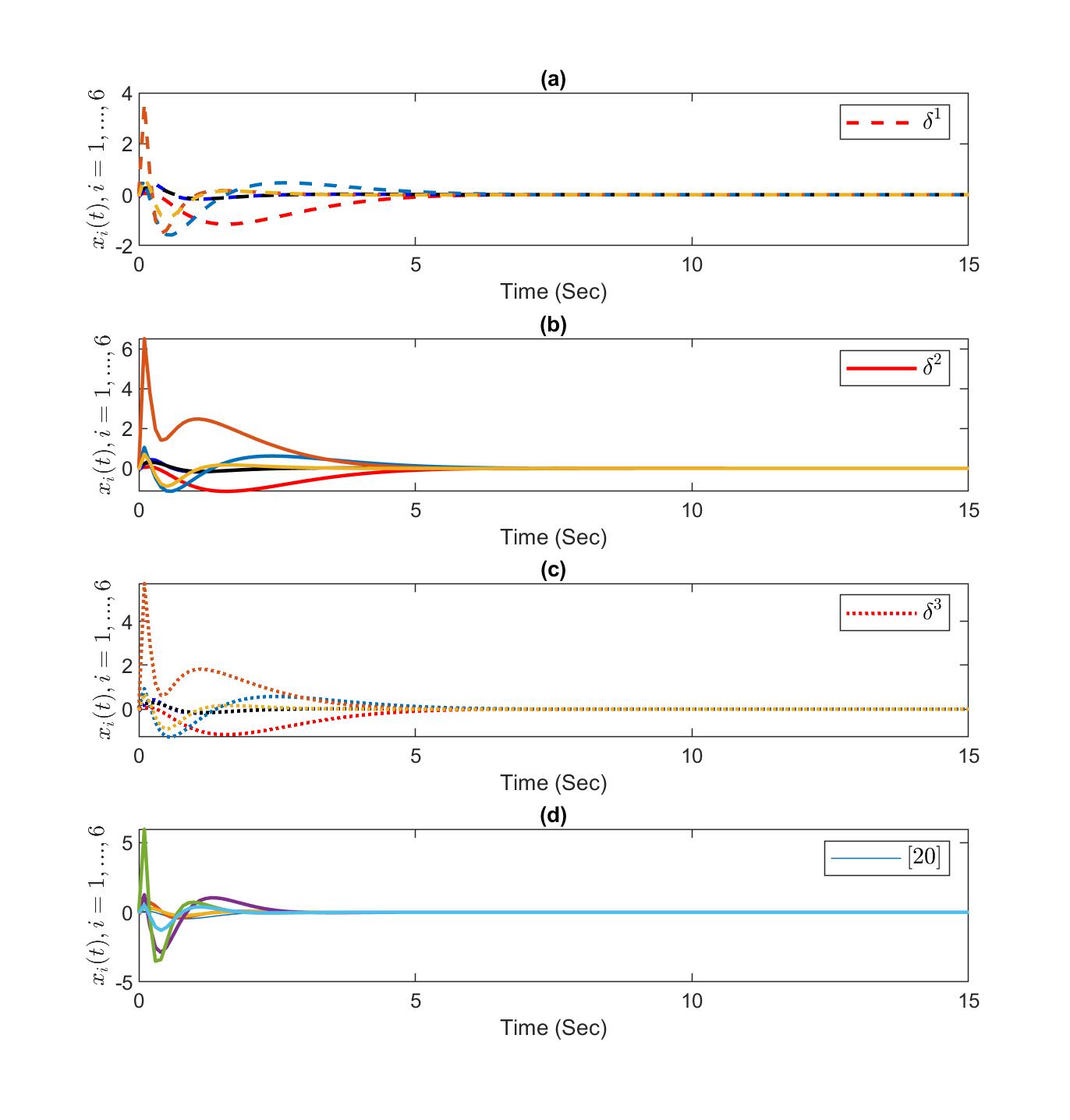}  
\caption{Comparison of the system state trajectories for three aspiration levels $\delta^i, i=1,2,3$ and method \cite{Lopez:2019}.} \label{Fig 1}
\end{center}
\end{figure}

\begin{figure}
\begin{center}
\includegraphics[width=7.6cm]{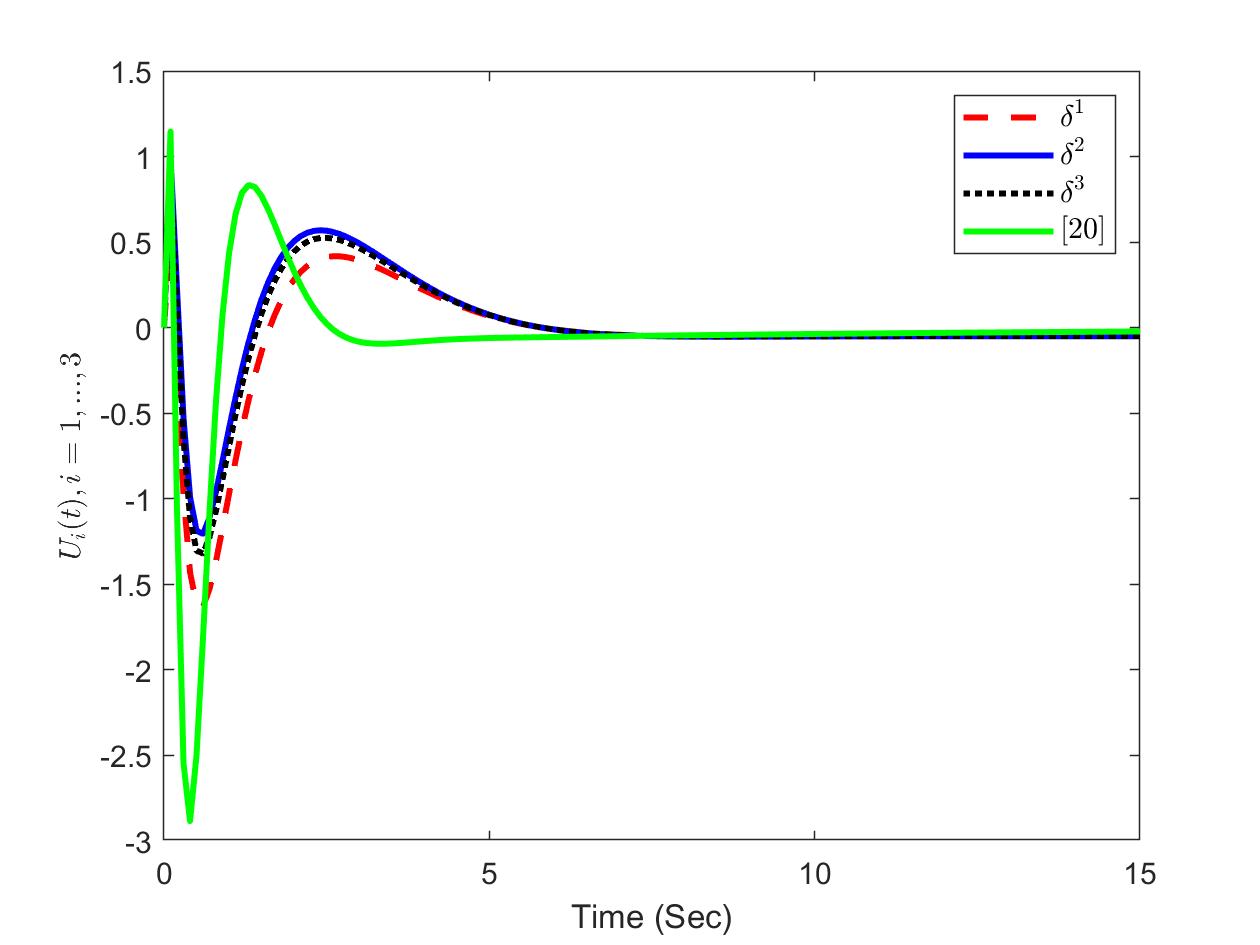}  
\caption{Comparison of the control policies for three aspiration levels $\delta^i, i=1,2,3$ and method \cite{Lopez:2019}.} \label{Fig 2}
\end{center}
\end{figure}

\textbf{Example 2}. Consider the quarter-suspension system depicted in Fig.~\ref{Fig 4}, with dynamics given by \cite{Jiang:2015} and \cite{Gaspar:2003}

\begin{figure}
\begin{center}
\includegraphics[width=7.3cm]{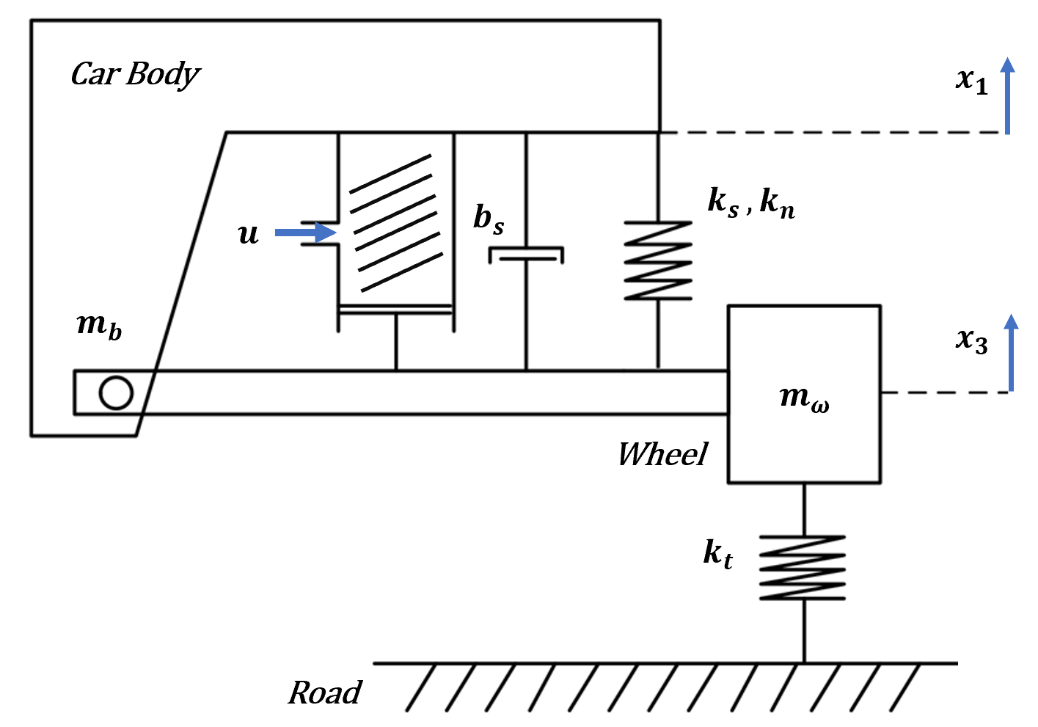}  
\caption{Quarter-car block diagram \cite{Zhu:2018}.} \label{Fig 4}
\end{center}
\end{figure}

\begin{align}
\begin{array}{l}
{{\dot x}_1} = {x_2}\\
{{\dot x}_2} =  - \frac{{{k_s}({x_1} - {x_3}) + {k_n}{{({x_1} - {x_3})}^3}}}{{{m_b}}}\, - \frac{{{b_s}({x_2} - {x_4}) - u}}{{{m_b}}}\\
{{\dot x}_3} = {x_4}\\
{{\dot x}_4} = \frac{{{k_s}({x_1} - {x_3}) + {k_n}{{({x_1} - {x_3})}^3}}}{{{m_\omega }}}\, + \frac{{{b_s}({x_2} - {x_4}) + {k_t}{x_3} - u}}{{{m_\omega }}}
\end{array}
\end{align}

\begin{figure}
\begin{center}
\includegraphics[width=9cm]{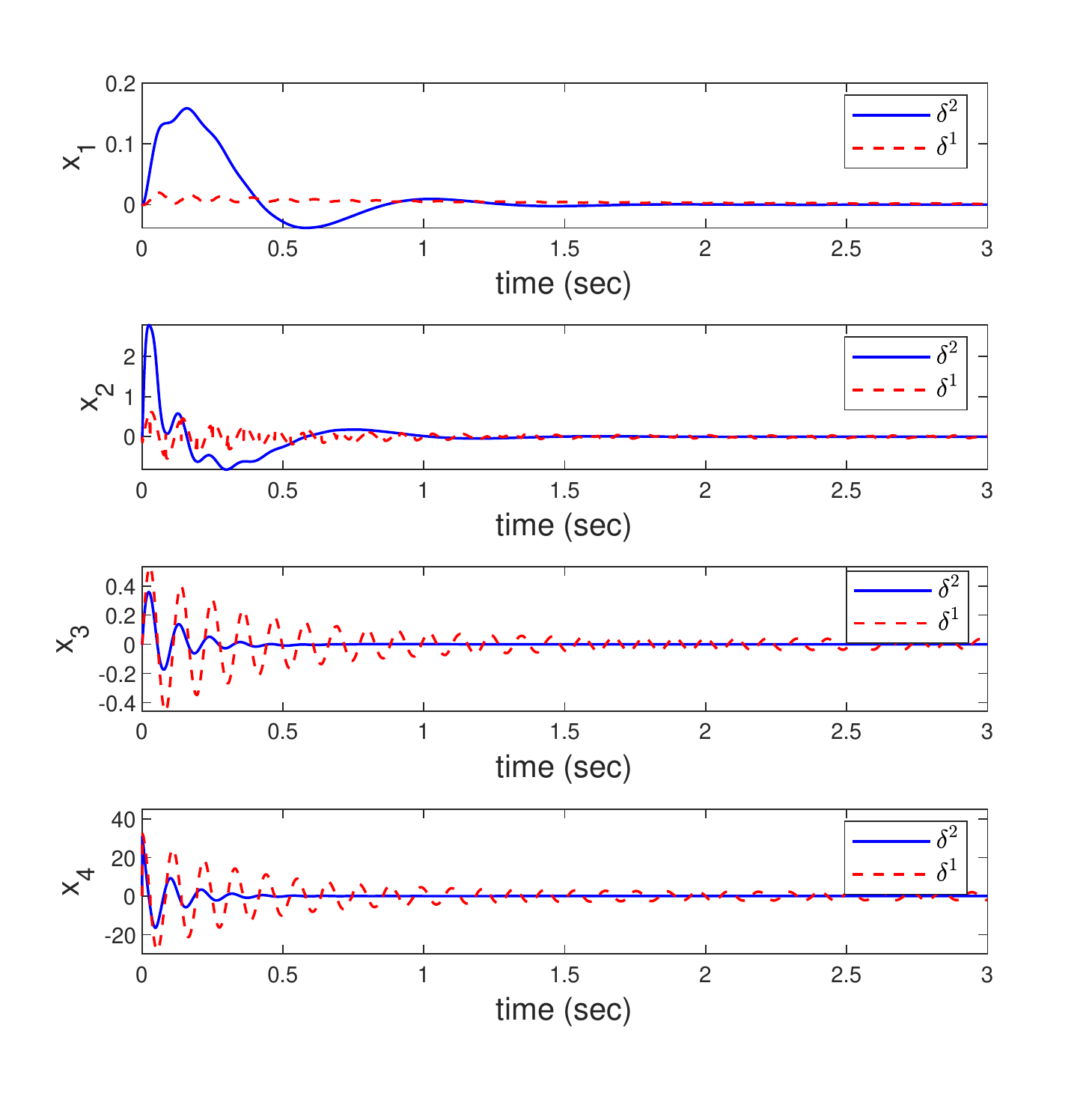}  
\caption{Comparison of performances with two learned control policies corresponding two different aspiration levels with no control policy.} \label{Fig 5}
\end{center}
\end{figure}
\noindent where ${x_1}$, ${x_2}$, ${x_3}$, and ${x_4}$ represent the position and velocity of the car and the position and velocity of the wheel assembly, respectively; ${m_b}$ and ${m_\omega }$  denote the mass of the car and the mass of the wheel assembly; ${k_t}$, ${k_s}$, ${k_n}$, and  ${b_s}$ denote the tyre stiffness, the linear suspension stiffness, the nonlinear suspension stiffness, and the damping rate of the suspension. Moreover, 
 $u$ is the control force from the hydraulic actuator.

Let ${m_b} = 300kg$, ${m_\omega } = 60kg$, ${k_t} = 190000N/m$, ${k_s} = 16000N/m$, ${k_n} = 1600$, and ${b_s} = 1000N/m/s$. The proposed data-driven algorithm is utilized to design control scheme for an active suspension system which simultaneously optimizes the following cost functions
\begin{align}
{J_1}({x_0},u) = \int_0^\infty  {(\sum\limits_{i = 1}^2 {10x_i^2 + {u^2}} )dt} 
\end{align}
\begin{align}
{J_2}({x_0},u) = \int_0^\infty  {(\sum\limits_{i = 3}^4 {x_i^2 + {u^2}} )dt}
\end{align}
in the set $\Omega  = \{ x\left| {x \in {\Re^4}and} \right.\,\left| {{x_1}} \right| \le 0.05,\,\,\left| {{x_2}} \right| \le 5,\,\left| {{x_3}} \right| \le 0.05,\,\,\left| {{x_4}} \right| \le 10\,\}$. Note that the main aim here is to maintain position and velocity of the car body, i.e., ${x_1}$ and ${x_2}$, as possible to maximize the passenger comfort while having satisficing performance on the position and velocity of the wheel assembly, i.e., ${x_3}$, and ${x_4}$, to reduce the fatigue of the quarter-suspension system.

Let us choose products of the set $\left\{ {{x_1},{\rm{ }}{x_2},{\rm{ }}{x_3},{\rm{ }}{x_4},{\rm{ }}x_1^2,{\rm{ }}x_2^2,{\rm{ }}x_3^2,{\rm{ }}x_4^2} \right\}$ with itself as the monomials for the value functions. One can see that the system is stable at the origin without any control input, so, the initial controller is chosen as ${u^0}(x) = 0$. We choose two different aspiration levels as

\vspace{-.2in}

 \begin{align}
 \begin{array}{l}
 {\delta ^1} = 10(0.3x_1^2{x_3} + 0.3x_1^2{x_4} + 0.1x_1^2 + 0.1{x_1}{x_2}{x_3} \\
  + 0.14{x_1}{x_2}+ 0.3x_2^2{x_3} + 0.1x_2^2 + 0.2{x_2}x_3^2 + 0.5{x_2}{x_3}{x_4} \\
+ 0.2{x_1}{x_3}{x_4} + 0.1{x_2}{x_4} + 4.12x_3^3 + 0.3x_3^2{x_4} + 0.47{x_3}x_4^2)
 \end{array}
 \end{align}
 \vspace{-.2in}
 \begin{align}
 \begin{array}{l}
 {\delta ^2} = 0.1(0.3x_1^2{x_3} + 0.3x_1^2{x_4} + 0.1x_1^2 + 0.1{x_1}{x_2}{x_3} \\
 + 0.14{x_1}{x_2}+ 0.3x_2^2{x_3} + 0.1x_2^2 + 0.2{x_2}x_3^2 + 0.5{x_2}{x_3}{x_4} \\
+ 0.2{x_1}{x_3}{x_4}+ 0.1{x_2}{x_4} + 4.12x_3^3 + 0.3x_3^2{x_4} + 0.47{x_3}x_4^2)
 \end{array}
 \end{align}
After the implementation of Algorithm 2 with two different aspiration levels, two suboptimal control policies are obtained, after seven and four iterations, as follows
\begin{align}
\begin{array}{l}
u_{{\delta ^1}}^{(7)}(x) =  - 11.94x_1^3 - 17.05x_1^2{x_2} + 5.44x_1^2{x_3} + 0.30x_1^2{x_4} \\
 + 0.00035x_1^2 - 11.06{x_1}x_2^2+ 11.95{x_1}{x_2}{x_3} - 0.62{x_1}{x_2}{x_4} \\
+ 0.00014{x_1}{x_2} - 33.95{x_1}x_3^2 + 3.37{x_1}{x_3}{x_4}- 0.00024{x_1}{x_3}  \\
- 2.80{x_1}x_4^2- 0.000046{x_1}{x_4} - 18.64{x_1} - 3.36x_2^3 + 8.51x_2^2{x_3} \\
- 1.34x_2^2{x_4} + 0.000059x_2^2+ 4.52{x_2}x_3^2 + 1.52{x_2}{x_3}{x_4}  \\
- 0.00014{x_2}{x_3} - 1.61{x_2}x_4^2 + 0.000062{x_2}{x_4} - 27.66{x_2} \\
+ 41.72x_3^3 + 0.34x_3^2{x_4} - 0.00021x_3^2 + 4.47{x_3}x_4^2+ 0.00010{x_3}{x_4} \\
+ 12.73{x_3} - 0.014x_4^3 - 0.0000083x_4^2 + 0.31{x_4}
\end{array}
\end{align}

\begin{align}
\begin{array}{l}
u_{{\delta ^2}}^{(4)} =  - 0.51x_1^3 - 0.13x_1^2{x_2} + 0.23x_1^2{x_3} + 0.026x_1^2{x_4}\\
 + 0.00000000017x_1^2 - 0.014{x_1}x_2^2 + 0.079{x_1}{x_2}{x_3}\\
 + 0.00135{x_1}{x_2}{x_4} + 0.00000000029{x_1}{x_2} - 0.502{x_1}x_3^2\\
 - 0.048{x_1}{x_3}{x_4} - 0.00000000131{x_1}{x_3} - 0.00508{x_1}x_4^2\\
 + 0.0000000000995{x_1}{x_4} - 0.361{x_1} + 0.00077x_2^3\\
 + 0.0192x_2^2{x_3} - 0.00031x_2^2{x_4} + 0.00000000000035x_2^2\\
 - 0.0277{x_2}x_3^2 + 0.0043{x_2}{x_3}{x_4} - 0.000000000038{x_2}{x_3}\\
 - 0.00039{x_2}x_4^2 - 0.000000000016{x_2}{x_4} - 0.114{x_2}\\
 + 0.775x_3^3 + 0.114x_3^2{x_4} + 0.0000000030x_3^2 + 0.025{x_3}x_4^2\\
 + 0.000000000031{x_3}{x_4} + 0.38{x_3} + 0.000109x_4^3\\
 + 0.00000000000092x_4^2 + 0.00030{x_4}
\end{array}
\end{align}

To test the learned controllers, a disturbance as a single pulse bump with the magnitude of $10$ is simulated at $t = [0 \sim 0.001]$ sec  such that the states deviate from the origin. The trajectories of the states after applying $u_{{\delta ^1}}^{(7)}(x)$  and  $u_{{\delta ^2}}^{(4)}$ are given in Fig.~\ref{Fig 5}. For the aspiration level ${\delta ^1}(x)$, the level of optimality on the second performance objective is not tight, therefore, we learn controller that has a better performance on the first performance objective, i.e., ${x_1}$ and ${x_2}$. For the aspiration level ${\delta ^2}(x)$, however, the level of optimality on the second performance objective is tighter. Therefore, the learned control policy has a better performance on the second performance objective.

\section{Conclusion}
This paper has developed an iterative data-driven adaptive dynamic programming algorithm for dynamic MO optimal control problem for nonlinear continues-time polynomial systems. The MO optimal control problem was, first, formulated as a aspiration-satisfying optimization problem with HJB inequalities as constraints. To deal with this problem, then, a 
MO SOS based iterative algorithm was presented to find some Pareto optimal solutions of MO optimal control problem with HJB inequalities. This MO SOS based iterative algorithm required the knowledge of the system dynamic. To obviate the requirement of complete knowledge of the system dynamics, an online data-driven reinforcement learning method was proposed for online implementation of the proposed MO SOS based algorithm. Finally, two simulation examples were provided to show the effectiveness of the proposed algorithm.


%

\appendices




\ifCLASSOPTIONcaptionsoff
  \newpage
\fi


\begin{thebibliography}{1}


\bibitem {Toivonen:1986}
Toivonen, H. T. (1986). A primal‐dual method for linear‐quadratic gaussian control problems with quadratic constraints. Optim. Control Appl. Methods, 7 (3), 305–314.

\bibitem {Toivonen:1989}
Toivonen, H. T., and Makila, P. M. (1989). Computer-aided design procedure for multiobjective LQG control problems. Int. J. Control, 49 (2), 655–666.

\bibitem {Marler:2004}
Marler, R. T., and Arora, J. S. (2004). Survey of multi-objective optimization methods for engineering. Structural and Multidisciplinary Optimization, 26 (6), 369–395. 

 \bibitem {Gambier:2011}
 Gambier, A.  and Jipp, M. (2011). Multi-objective optimal control: An introduction. ASCC 2011 - 8th Asian Control Conf. - Final Progr. Proc., 1084–1089.
 
 \bibitem {Peitz:2018}
Peitz, S. and Dellnitz, M. (2018).  A Survey of Recent Trends in Multi-objective Optimal Control—Surrogate Models, Feedback Control and Objective Reduction. Math. Comput. Appl., 23(2), 30.  
 
 
 \bibitem {Roijers:2013}
 Roijers, D. M., Vamplew, P., Dazeley, R., Whiteson, S. and Dazeley, R. (2013). A survey of multi-objective sequential decision-making. J. Artif. Intell. Res., 48, 67–113.
 
 
 
 \bibitem {Logist:2010}
 Logist, F., Sager, S., Kirches, C., and Van Impe, J. F. (2010). Efficient multiple objective optimal control of dynamic systems with integer controls. J. Process Control, 20 (7), 810–822.
 
 \bibitem {Ober-Blobaum:2012}
 Ober-Blobaum, S., Ringkamp, M., and Zum Felde, G. (2012). Solving multiobjective Optimal Control problems in space mission design using Discrete Mechanics and reference point techniques. Proc. IEEE Conf. Decis. Control, 5711–5716.
 
 \bibitem {Caramia:2008}
 Caramia, M., and Dell'Olmo, P. (2008). Multi-objective optimization. in Multi-Objective Management in Freight Logistics. Increasing Capacity, Service Level and Safety with Optimization Algorithms. London, U.K. Springer.
 
 \bibitem {Das:1997}
 Das, I., and Dennis, J. E. (1997). A closer look at drawbacks of minimizing weighted sums of objectives for Pareto set generation in multicriteria optimization problems. Struct. Optim., 14 (1), 63–69.
 
 
 \bibitem {Lewis:2009}
 Lewis, Frank L., and Draguna Vrabie. (2009). Adaptive Dynamic Programming for Feedback Control. Proceedings of 2009 7th Asian Control Conference, ASCC 2009, 1402–9.
 
 \bibitem {Vamvoudakis:2010}
 Vamvoudakis, K. G.  and Lewis, F. L. (2010). Online actor-critic algorithm to solve the continuous-time infinite horizon optimal control problem. Automatica, 46(5), 878–888. 
 
 
  \bibitem {Jiang:2015}
 Jiang, Y., and Jiang, Z.  (2015). Global Adaptive Dynamic Programming for Continuous-Time Nonlinear Systems. in IEEE Transactions on Automatic Control, 60(11). 2917-2929.
 
  \bibitem {Wang:2014}
  Wang, Y., O'Donoghue, B., and Boyd, S. (2014). Approximate dynamic programming via iterated Bellman inequalities. International Journal of Robust and Nonlinear Control, 25(10), 1472–1496. 
 
 \bibitem {Kamalapurkar:2018}
 Kamalapurkar, R., Walters, P., Rosenfeld, J., and Dixon, W. (2018). Reinforcement Learning for Optimal Feedback Control. Cham: Springer International Publishing.
 
 
 \bibitem {Modares:2016}
 Modares, H.,  Lewis, F. L., and Jiang, Z. P.  (2016). Optimal Output-Feedback Control of Unknown Continuous-Time Linear Systems Using Off-policy Reinforcement Learning. IEEE Trans. Cybern., 46(11), 2401–2410.
 
 

 
 
 \bibitem {Beuchat:2020}
 Beuchat, P. N., Georghiou, A., and Lygeros, J. (2020). Performance Guarantees for Model-Based Approximate Dynamic Programming in Continuous Spaces. IEEE Transactions on Automatic Control, 65(1), 143–158.
 
 
 \bibitem {Tanzanakis:2020}
 Tanzanakis, A., and Lygeros, J. (2020). Data-Driven Control of Unknown Systems: A Linear Programming Approach. ArXiv ID:2003.00779.
 
 
 \bibitem {Kang:2004}
 Kang, D. O., and Bien, Z. (2004). Multi-objective control problems by reinforcement learning. in Handbook of Learning and Approximate Dynamic Programming, 433–461.
 
  \bibitem {Lopez:2019}
 Lopez, V. G., and Lewis, F. L. (2019). Dynamic Multi-objective Control for Continuous-Time Systems Using Reinforcement Learning. IEEE Trans. Automat. Contr., 64(7), 2869–2874.
 
 \bibitem {Moffaert:2014}
 Moffaert, K. V., and Now{{\'e}}, A. (2014). Multi-Objective Reinforcement Learning using Sets of Pareto Dominating Policies. Journal of Machine Learning Research, 15(107), 3663--3692.
 
 
  \bibitem{Barrett:2008}
 Barrett, L., and Narayanan, S. (2008). Learning all optimal policies with multiple criteria. ICML '08.
 
 \bibitem{Chen:2019}
  Chen, X.,  Ghadirzadeh, A.,  Björkman, M., and  Jensfelt, P. (2019). Meta-Learning for Multi-objective Reinforcement Learning. 2019 IEEE/RSJ International Conference on Intelligent Robots and Systems (IROS), Macau, China, 977--983.
  
  
   
  \bibitem{Abouheaf:2019} 
   Abouheaf, M., Gueaieb, W., and Spinello, D. (2019). Online Multi-Objective Model-Independent Adaptive Tracking Mechanism for Dynamical Systems. Robotics, 8, 82.
  
  
  \bibitem {Carmichael:1980}
  Carmichael, D. G. (1980). Computation of Pareto optima in structural design. Int. J. Numer. Methods Eng., 15 (6), 925–929.
  
  
  
   \bibitem{Ursem:2009} 
   Ursem, R. K. (2009). Multi-objective Optimization using Evolutionary Algorithms.
  
   
    
   \bibitem {Lewis:2012}
   Lewis, F.  L., Vrabie, D. L., and Syrmos, V. L. (2012). Optimal Control: Third Edition. Hoboken, NJ, USA: John Wiley and Sons, Inc.
   

   
     
\bibitem {Ahmadi:2018}
Ahmadi, A. (2018). Sum of Squares (SOS) Techniques: An Introduction. 1--9.



\bibitem {Li:2008}
 Li, Q. R.,  Tao, W. H.,  Sun, N.,  Zhang, C. Y., and  Yao, L. H. (2008). Stabilization control of double inverted pendulum system. In Proc. 3rd Int. Conf. Innovative Comput. Inf. Control, Jun. 18--20.
 
 \bibitem {Gaspar:2003}
  Gaspar, P.,  Szaszi, I., and  Bokor, J. (2003). Active suspension design using linear parameter varying control. Int. J. Veh. Auton. Syst., 1(2), 206-221. 
 
 \bibitem {Zhu:2018}
  Zhu, Y., Zhao, D., Yang, X., and Zhang, Q. (2018). Policy Iteration for $H_\infty $ Optimal Control of Polynomial Nonlinear Systems via Sum of Squares Programming. IEEE Transactions on Cybernetics, 48(2), 500–509.
 
 
 

















\end{thebibliography}
\end{document}